\documentclass[showkeys,showpacs,amsmath,amssymb,a4paper,12pt]{revtex4}
\usepackage{amsmath}
\usepackage{amssymb}
\usepackage{amsmath,amsthm,amssymb,amscd}
\usepackage{latexsym}
\usepackage{indentfirst}
\usepackage{subfigure}
\usepackage{graphicx}
\usepackage[top=1in,bottom=1in,left=1.25in,right=1.25in]{geometry}

\begin{document}

\title{
  \bf Totally asymmetric simple exclusion process with one or two shortcuts
}

\author{Rui Gu}
\email{gurui88@sina.com }
\affiliation{ School of Mathematical Sciences, Fudan University,
Shanghai 200433, China}

\author{Yunxin Zhang}
\email{xyz@fudan.edu.cn}
\affiliation{ School of Mathematical Sciences and
Centre for Computational System Biology, Fudan University, Shanghai 200433, China}

\begin{abstract}
In this paper, the operation of totally asymmetric simple exclusion
process with one or two shortcuts under open boundary conditions is
discussed. Using both mathematical analysis and numerical
simulations, we have found that, according to the method chosen by
the particle at the bifurcation, the model can be separated into two
different situations which lead to different results. The results obtained in this paper would be very useful
in the road building, especially at the bifurcation of the road.
\end{abstract}



\maketitle

\baselineskip=22pt

\section{Introduction}
One-dimensional totally asymmetric simple exclusion process (TASEP)
has been studied for many years \cite{1, a}. It is one special
type of ASEPs which represent one of the basic models studying
non-equilibrium behavior of particle transport along one-dimensional
lattices. ASEP was first introduced in \cite{2} for the explanation
of ribosome inside the cell of creature. Now, ASEP has also been used to
simulate a lot of physical processes including surface diffusion
\cite{3}, traffic model \cite{4}, and molecular motors \cite{b},\cite{c},
etc.

TASEP, as a special case of ASEP, has two identical characters. One
is that the model is discrete, which means that it is studied in
finite time intervals. The other is that the particles in the
lattices can only move in one direction. The analytical solutions
under open boundary conditions had been obtained in \cite{5, 6}.

Under open boundary conditions, the solutions yield phase diagrams
with three phases \cite{5,6}. At small values of injection rates
$\alpha<0.5$ and $\alpha<\beta$, the system is found in a
low-density entry-limited phase where
\begin{equation}
\rho_{1}=\alpha \quad
\quad\rho_{L}=\frac{\alpha(1-\alpha)}{\beta}\quad\quad
J=\alpha(1-\alpha) \quad \quad\rho_{bulk}=\alpha\\
\end{equation}
where $\rho_{1},\rho_{L}$ and $\rho_{bulk}$ are the densities at the
entrance, exit and the bulk of the lattice far away from the
boundaries, respectively. $J$ denotes the flux.

At small values of extraction rates $\beta < 0.5$ and $\beta <
\alpha$, the system is in a high-density exit-limited phase with\\
\begin{equation}
\rho_{1}=1-\frac{\beta(1-\beta)}{\alpha}\quad\quad\rho_{L}=1-\beta
\quad\quad J=\beta(1-\beta)\quad\quad\rho_{bulk}=1-\beta\\
\end{equation}
At large values of the injection ($\alpha\leqslant0.5$) and
extraction ($\beta\leqslant0.5$) rates the system is in a
maximal current phase with\\
\begin{equation}
\rho_{1}=1-\frac{1}{4\alpha}\quad\quad\rho_{L}=\frac{1}{4\beta}
\quad\quad J=\frac{1}{4}\quad\quad \rho_{bulk}=\frac{1}{2}\\
\end{equation}
A large number of varieties of extensions of TASEPs have been
investigated, such as TASEP with hierarchical long-range connections
\cite{7}, two lane situations \cite{8}, two speed TASEP\cite{8.1},
the effect of defect locations\cite{9}, particle-dependent hopping
rates \cite{10}, and so forth.

Recently, J. Brankov \cite{11} and E. Pronina \cite{12} studied an
ASEP with two chains in the middle of the filament. They supposed
that a particle chooses to move into these two chains with equal
probability, 0.5 and these two chains have the same length. Then in
2007, Yao-Ming Yuan and Rui Jiang \cite{13} investigated a TASEP
with a shortcut in the middle. They set a possibility $q$ for a
particle to jump through the shortcut when it faces the bifurcation,
and they set the length of the shortcut to be zero. Unfortunately,
they did not
state the situation clearly.

In this paper, we will investigate TASEP with one or two
shortcuts respectively. The
length of the shortcut is also assumed to be zero. For example, the filament on which the
motor moves may be twisted as figure 1(a), a motor may
have a chance to jump directly from site $k_{1}$ to $k_{2}$, as shown in
figure 1(b).

To the basic model, i.e. there is only one shortcut along the filament.
It is natural for us to divide the whole filament into three
segments. As shown in figure 1(b), a molecular motor will face a
choice of whether to jump through the shortcut or to move ordinarily
through segment 2, when it reaches site $k_{1}$. An important
problem is that the particle at site $k_{1}$ may have to wait to go
through the shortcut if there is also another particle occupying
site
$k_{2}-1$. Which of them to go first should be determined clearly.

After the basic model investigated, we would like to do some
research work on our advanced model 1. In advanced model 1, there
are two shortcuts which begin and end at different sites. These two
shortcuts stay respectively along the filament, that is to say,
shortcut 2 begins after shortcut 1 ends, as shown in figure 2.

Also, considering the difference of places where the shortcuts stay,
we have advanced model 2, as shown in figure 3. In advanced model 2
there are also two shortcuts. But they begin at the same site and
end at different, which means when a motor moves into site $k_{1}$,
it will choose whether to pass through shortcut 1 or
shortcut 2 or to go ordinarily through segment 2.

In the following, we firstly give the detailed description of the different models, then we will discuss their phase situation of the corresponding segments theoretically, which is followed by the numerical simulations.
\begin{figure}
\begin{minipage}[t]{1\linewidth}
    \centering
    \includegraphics[width=4in]{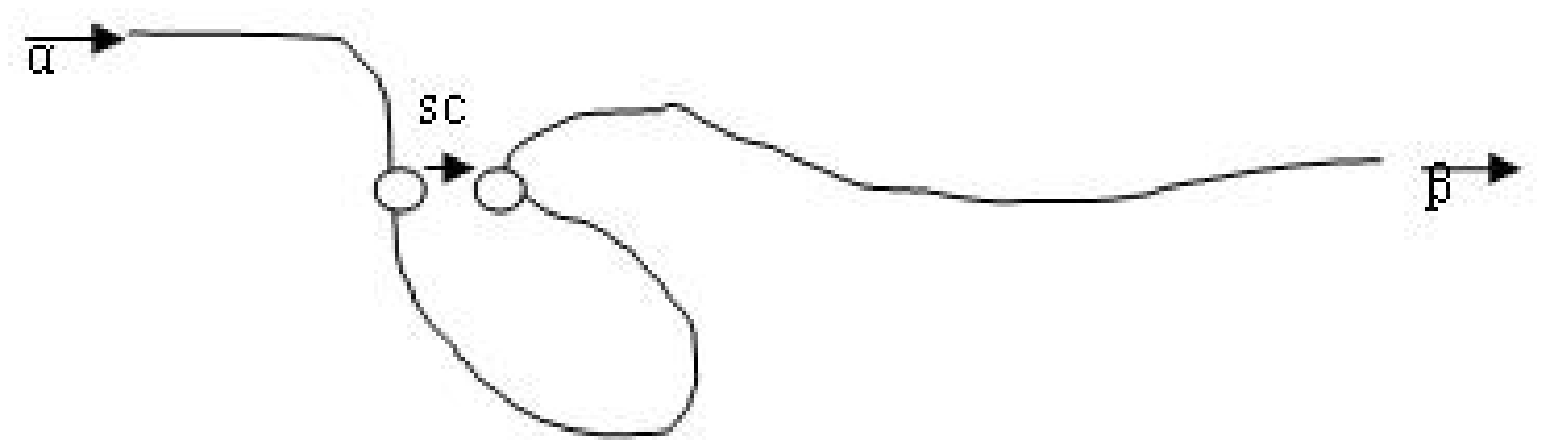}
    \label{fig:side:a}
    \subfigure{\footnotesize (a)}
  \end{minipage}
  \begin{minipage}[t]{1\linewidth}
    \centering
    \includegraphics[width=4in]{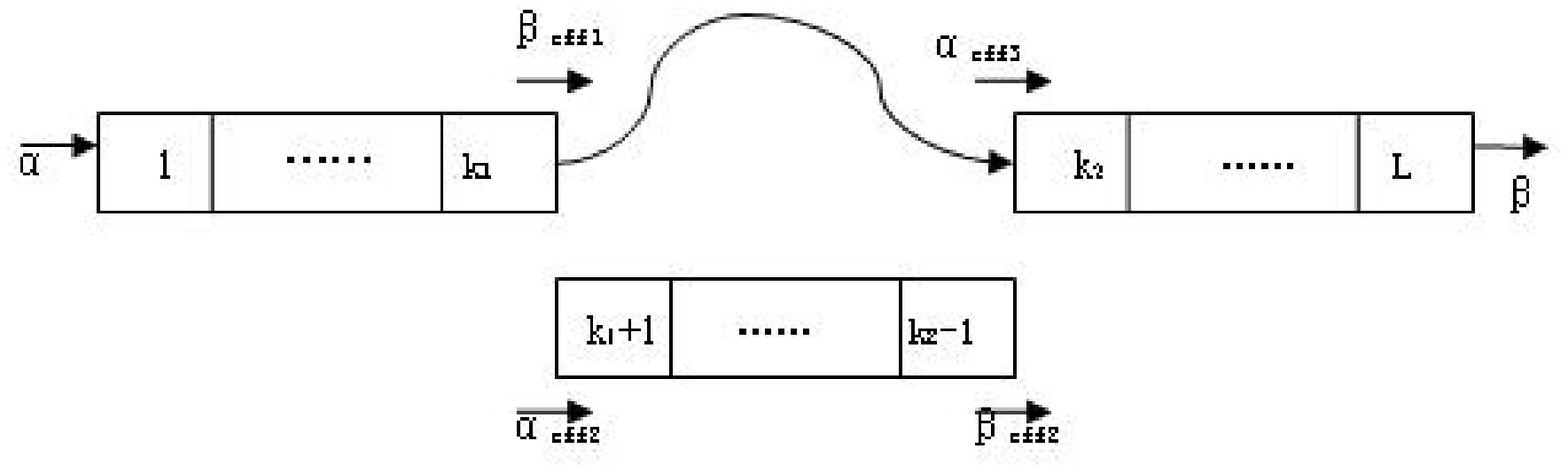}
    \label{fig:side:b}
    \subfigure{\footnotesize (b)}
  \end{minipage}
\caption{\footnotesize basic model with one shortcut: (a)a road with
one shortcut at the bulk; (b)lattices with one shortcut.}
\end{figure}
\begin{figure}
\begin{minipage}[t]{1\linewidth}
    \centering
    \includegraphics[width=4in]{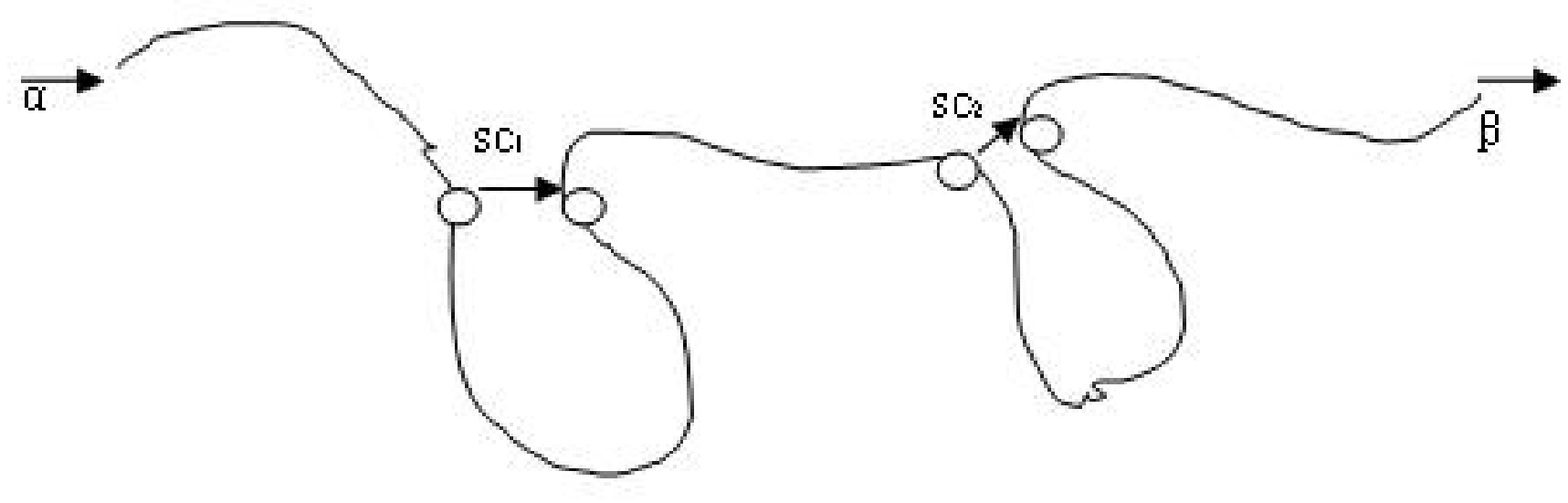}
    \label{fig:side:a}
    \subfigure{\footnotesize (a)}
  \end{minipage}
  \begin{minipage}[t]{1\linewidth}
    \centering
    \includegraphics[width=5in]{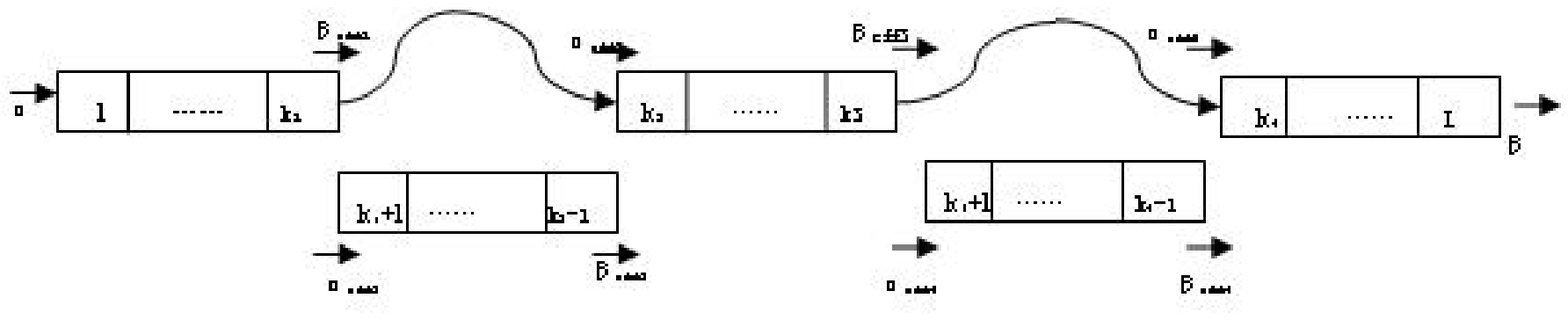}
    \label{fig:side:b}
    \subfigure{\footnotesize (b)}
  \end{minipage}
\caption{\footnotesize advanced model with two shortcuts at
different beginnings and endings: (a)a road with two shortcuts at
different beginnings and endings; (b)lattices with two shortcuts at
different beginnings and endings.}
\end{figure}
\begin{figure}
\begin{minipage}[t]{1\linewidth}
    \centering
    \includegraphics[width=4in]{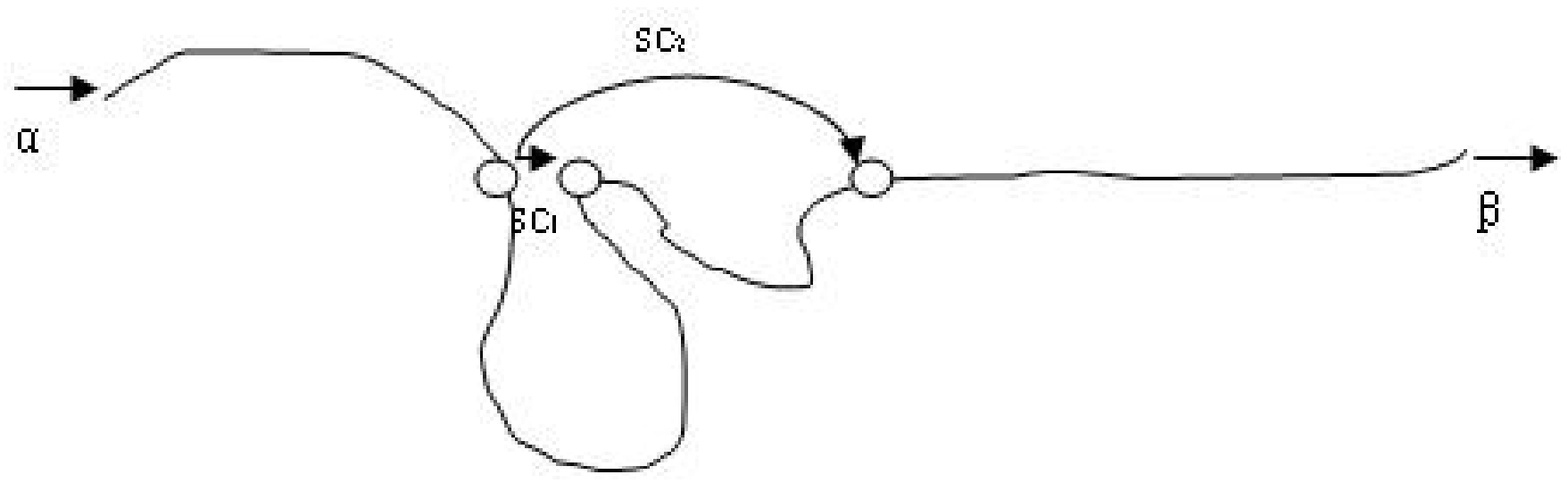}
    \label{fig:side:a}
    \subfigure{\footnotesize (a)}
  \end{minipage}
  \begin{minipage}[t]{1\linewidth}
    \centering
    \includegraphics[width=4in]{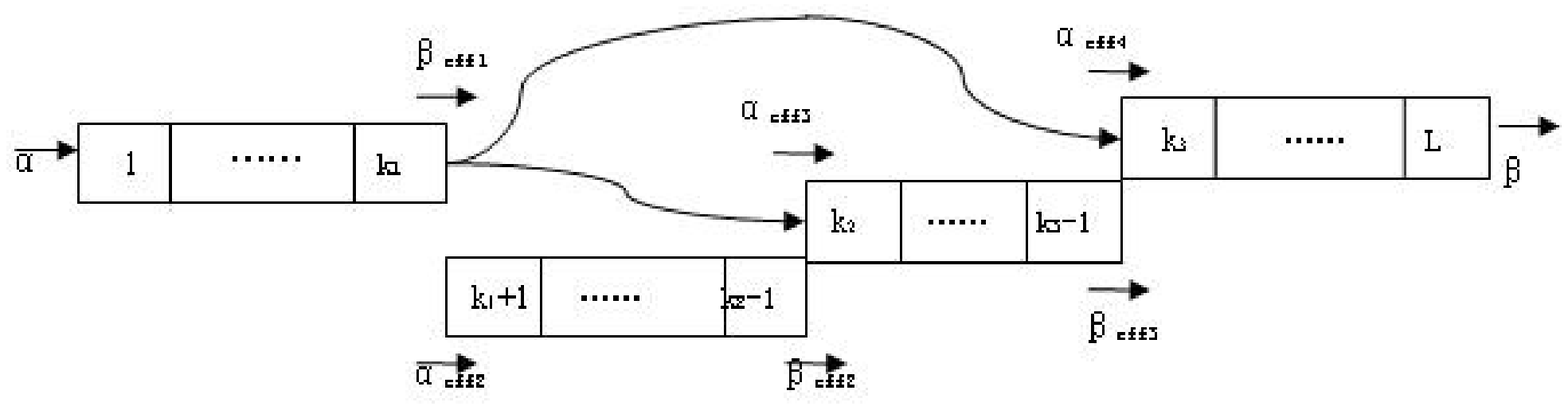}
    \label{fig:side:b}
    \subfigure{\footnotesize (b)}
  \end{minipage}
\caption{\footnotesize advanced model with two shortcuts at the same
beginning and different endings: (a)a road with two shortcuts at the
same beginning and different endings; (b)lattices with two shortcuts
at the same beginning and different endings.}
\end{figure}

\section{Models}
\subsection{Basic model}
First, we divide the lattices into three segments, as shown in
figure 1(b). Segment 1,2,3 start from site $1,k_{1}+1,k_{2}$ and
end in site $k_{1},k_{2}-1,L$, respectively.

In this model, all the particles traveling in the lattices are
identical to each other. At time $t=0$, we may suppose that the
lattices are empty. As time goes, there will be a particle in
site 1. This particle will pass continuously through the whole
filament.

Given a site $i(1\leqslant i\leqslant L)$, in an infinitesimal
time interval $dt$, if $i=1$, a particle is inserted with
probability $\alpha dt$, provided the site is empty, if $i=L$ and
there is a particle in it, then the particle in site L is extracted
with probability $\beta dt$, and if $1\leqslant i\leqslant L(i\neq
k_{1},k_{2}-1)$ and site $i$ is occupied, the particle in site $i$
will move into site $i+1$ with probability $dt$, providing site
$i+1$ is empty. However, if a particle is in site $k_{1}$ or
$k_{2}-1$, we have to clarify the way it moves. Although Yao-Ming
Yuan,Rui Jiang \cite{13} have discussed the method of choice at site
$k_{1}$, the situation at site $k_{2}-1$ has not been studied
clearly. It is possible that when both site $k_{1}$ and site
$k_{2}-1$ have particles, it may produce a collision
in site $k_{2}$.

\subsubsection{Situation 1}
If site $k_{1}$ has a particle, then the particle will choose the
way
as following:

$\bullet$ if site $k_{1}+1$ and site $k_{2}$ are both occupied, then
the
particle in site $k_{1}$ does not move.

$\bullet$ if site $k_{2}$ is occupied but site $k_{1}+1$ is empty,
then the particle in site $k_{1}$ will move into site $k_{1}+1$ with
probability $dt$.

$\bullet$ if site $k_{1}+1$ is occupied but site $k_{2}$ is empty then:

\indent$\star$ if site $k_{2}-1$ is empty, the particle in site
$k_{1}$ will move into site $k_{2}$ with probability $qdt$.

\indent$\star$ if site $k_{2}-1$ is occupied , the particle in
site $k_{1}$ will move into site $k_{2}$ with probability $qpdt$.

$\bullet$ if both site $k_{1}+1$ and site $k_{2}$ are empty, then
the particle in site $k_{1}$ will move into site $k_{1}+1$ with
probability
$(1-q)dt$, and:

\indent$\star$ if site $k_{2}-1$ is empty, the particle in site
$k_{1}$
will move into site $k_{2}$ with probability $qdt$.

\indent$\star$ if site $k_{2}-1$ is occupied , the particle in site
$k_{1}$ will move into site $k_{2}$ with probability $qpdt$.

If site $k_{2}-1$ has a particle, and site $k_{2}$ is empty, then
the particle will choose the way
as following:

$\bullet$ if site $k_{1}$ is empty, then the particle in site
$k_{2}-1$
will move into site $k_{2}$ with probability $dt$.\\
$\bullet$ if site $k_{1}$ is occupied, then the particle in site
$k_{2}-1$ will move into site $k_{2}$ with probability $(1-pq)dt$.\\

The character of this model is that the particle in site $k_{1}$
first choose whether to move through the shortcut and once it has
decided to jump through the shortcut, it will face the problem of
whether site $k_{2}-1$ is occupied. If site $k_{2}-1$ is occupied,
then the particle which has chosen the shortcut can only move into
site $k_{2}$ with probability $pdt$ in order to prevent collision.
This means that the particle takes a risk to jump the shortcut
because once it has chosen the shortcut it can not go back into
segment 2, even if
site $k_{1}+1$ is empty.

\subsubsection{Situation 2}
If site $k_{1}$ has a particle, then the particle will choose the
way
as following:

$\bullet$ if site $k_{1}+1$ and site $k_{2}$ are both occupied, then
the
particle in site $k_{1}$ does not move.

$\bullet$ if site $k_{2}$ is occupied but site $k_{1}+1$ is empty,
then the particle in site $k_{1}$ will move into site $k_{1}+1$ with
probability $dt$.\\
$\bullet$ if site $k_{1}+1$ is occupied but site $k_{2}$ is empty then:

\indent$\star$ if site $k_{2}-1$ is empty, the particle in site
$k_{1}$ will move into site $k_{2}$ with probability $qdt$.

\indent$\star$ if site $k_{2}-1$ is occupied , the particle in
site $k_{1}$ will move into site $k_{2}$ with probability $qpdt$.

$\bullet$ if both site $k_{1}+1$ and site $k_{2}$ are empty, then:

\indent$\star$ if site $k_{2}-1$ is empty, the particle in site
$k_{1}$ will move into site $k_{2}$ with probability $qdt$, and it
will move
into site $k_{1}+1$ with probability $(1-q)dt$.

\indent$\star$ if site $k_{2}-1$ is occupied , the particle in site
$k_{1}$ will move into site $k_{2}$ with probability $qpdt$, and it
will
move into site $k_{1}+1$ with probability $(1-qp)dt$.

If site $k_{2}-1$ has a particle, and site $k_{2}$ is empty, then
the particle will choose the way
as following:

$\bullet$ if site $k_{1}$ is empty, then the particle in site
$k_{2}-1$
will move into site $k_{2}$ with probability $dt$.

$\bullet$ if site $k_{1}$ is occupied, then the particle in site
$k_{2}-1$
will move into site $k_{2}$ with probability $(1-pq)dt$.

The character of this model is that once the particle has chosen the
shortcut but see that the particle in site $k_{2}-1$ has priority to
move into site $k_{2}$, it can change its mind and return to site
$k_{1}+1$, providing site $k_{1}+1$ is empty. This model is closer
to reality because actually bus drivers can change their mind in
time when they are told that the road ahead is to some extent
blocked. It is not reasonable for drivers to take a shortcut which
is actually
difficult to go through.

\subsection{Advanced model 1}
Now we divide the lattices into five segments, as shown in figure
2(b),segment 1,2,3,4,5 starts in site
$1,k_{1}+1,k_{2},k_{3}+1,k_{4}$ and ends
in site $k_{1},k_{2}-1,k_{3},k_{4}-1,L$,respectively.

In this model, all the particles traveling in the lattices are
identical to each other. At time $t=0$, we may suppose that the
lattices are empty. As time goes, there will be a particle goes into
site 1. This particle will pass continuously through the whole
filament.

Given site $i(1\leqslant i\leqslant L)$, in an infinitesimal
time interval $dt$, if $i=1$, a particle is inserted with
probability $\alpha dt$, provided the site is empty, if $i=L$ and
there is a particle in it, then the particle in site L is extracted
with probability $\beta dt$, and if $1\leqslant i\leqslant L(i\neq
k_{1},k_{2}-1,k_{3},k_{4}-1)$ and site $i$ is occupied, the particle
in site $i$ will move into site $i+1$ with probability $dt$,
providing site $i+1$ is empty. However, if a particle is in site
$k_{1}$ or $k_{2}-1$ or $k_{3}$ or $k_{4}-1$, we have to clarify the
way it moves. We will construct advanced model 1 using similar
method according to basic model(situation 1).

If site $k_{1}$ has a particle, then the particle will choose the
way
as following:

$\bullet$ if site $k_{1}+1$ and site $k_{2}$ are both occupied, then
the
particle in site $k_{1}$ does not move.

$\bullet$ if site $k_{2}$ is occupied but site $k_{1}+1$ is empty,
then the particle in site $k_{1}$ will move into site $k_{1}+1$ with
probability $dt$.

$\bullet$ if site $k_{1}+1$ is occupied but site $k_{2}$ is empty then:

\indent$\star$ if site $k_{2}-1$ is empty, the particle in site
$k_{1}$ will move into site $k_{2}$ with probability $q_{1}dt$.

\indent$\star$ if site $k_{2}-1$ is occupied , the particle in
site $k_{1}$ will move into site $k_{2}$ with probability $q_{1}p_{1}dt$.

$\bullet$ if both site $k_{1}+1$ and site $k_{2}$ are empty, then
the particle in site $k_{1}$ will move into site $k_{1}+1$ with
probability
$(1-q_{1})dt$, and:

\indent$\star$ if site $k_{2}-1$ is empty, the particle in site
$k_{1}$
will move into site $k_{2}$ with probability $q_{1}dt$.

\indent$\star$ if site $k_{2}-1$ is occupied , the particle in site
$k_{1}$ will move into site $k_{2}$ with probability $q_{1}p_{1}dt$.

If site $k_{2}-1$ has a particle, and site $k_{2}$ is empty, then
the particle will choose the way
as following:

$\bullet$ if site $k_{1}$ is empty, then the particle in site
$k_{2}-1$
will move into site $k_{2}$ with probability $dt$.

$\bullet$ if site $k_{1}$ is occupied, then the particle in site
$k_{2}-1$
will move into site $k_{2}$ with probability $(1-p_{1}q_{1})dt$.

If site $k_{3}$ has a particle, then the particle will choose the
way
as following:

$\bullet$ if site $k_{3}+1$ and site $k_{4}$ are both occupied, then
the
particle in site $k_{3}$ does not move.

$\bullet$ if site $k_{4}$ is occupied but site $k_{3}+1$ is empty,
then the particle in site $k_{3}$ will move into site $k_{3}+1$ with
probability $dt$.

$\bullet$ if site $k_{3}+1$ is occupied but site $k_{4}$ is empty then:

\indent$\star$ if site $k_{4}-1$ is empty, the particle in site
$k_{3}$ will move into site $k_{4}$ with probability $q_{2}dt$.

\indent$\star$ if site $k_{4}-1$ is occupied , the particle in
site $k_{3}$ will move into site $k_{4}$ with probability $q_{2}p_{2}dt$.

$\bullet$ if both site $k_{3}+1$ and site $k_{4}$ are empty, then
the particle in site $k_{3}$ will move into site $k_{3}+1$ with
probability
$(1-q_{2})dt$, and:

\indent$\star$ if site $k_{4}-1$ is empty, the particle in site
$k_{3}$
will move into site $k_{4}$ with probability $q_{2}dt$.

\indent$\star$ if site $k_{4}-1$ is occupied , the particle in site
$k_{3}$ will move into site $k_{4}$ with probability $q_{2}p_{2}dt$.

If site $k_{4}-1$ has a particle, and site $k_{4}$ is empty, then
the particle will choose the way
as following:

$\bullet$ if site $k_{3}$ is empty, then the particle in site
$k_{4}-1$
will move into site $k_{4}$ with probability $dt$.

$\bullet$ if site $k_{3}$ is occupied, then the particle in site
$k_{4}-1$
will move into site $k_{4}$ with probability $(1-p_{2}q_{2})dt$.

\subsection{Advanced model 2}
Also we can divide the lattices into four segments, as shown in
figure 3(b), segment 1,2,3,4 starts in site $1,k_{1}+1,k_{2},k_{3}$
and ends
in site $k_{1},k_{2}-1,k_{3}-1,L$,respectively.

In this model, all the particles traveling in the lattices are
identical to each other. At time $t=0$, we may suppose that the
lattices are empty. As time goes, there will be particles goes into
site 1. These particles will pass continuously through the whole
filament.

Given site $i(1\leqslant i\leqslant L)$ , in an infinitesimal
time interval $dt$, if $i=1$, a particle is inserted with
probability $\alpha dt$, provided the site is empty, if $i=L$ and
there is a particle in it, then the particle in site L is extracted
with probability $\beta dt$, and if $1\leqslant i\leqslant L(i\neq
k_{1},k_{2}-1,k_{3}-1)$ and site $i$ is occupied, the particle in
site $i$ will move into site $i+1$ with probability $dt$, providing
site $i+1$ is empty. However, if a particle is in site $k_{1}$ or
$k_{2}-1$ or $k_{3}-1$, we have to clarify the way it moves. We will
construct advanced model 2 using similar
method according to basic model(situation 1).

If site $k_{1}$ has a particle, then the particle will choose the
way
as following:\\
$\bullet$ If sites $k_{1}+1$,$k_{2}$ and $k_{3}$ are all occupied,
the the particle in site $k_{1}$ does not move.

$\bullet$ If site $k_{1}+1$ is empty but sites $k_{2}$ and $k_{3}$
are occupied, the particle in site $k_{1}$ will move into site
$k_{1}+1$ with probability $dt$.

$\bullet$ If site $k_{2}$ is empty but sites $k_{1}+1$ and $k_{3}$
are occupied, then:

\indent$\star$ If site $k_{2}-1$ is empty, the particle in site
$k_{1}$ will move into site $k_{2}$ with probability $q_{1}dt$.

\indent$\star$ If site $k_{2}-1$ is occupied, the particle in site
$k_{1}$ will move into site $k_{2}$ with probability $q_{1}p_{1}dt$.

$\bullet$ If site $k_{3}$ is empty but sites $k_{1}+1$ and $k_{2}$
are occupied, then:

\indent$\star$ If site $k_{3}-1$ is empty, the particle in site
$k_{1}$ will move into site $k_{3}$ with probability $q_{2}dt$.

\indent$\star$ If site $k_{3}-1$ is occupied, the particle in site
$k_{1}$ will move into site $k_{3}$ with probability $q_{2}p_{2}dt$.

$\bullet$ If site $k_{1}+1$ and site $k_{2}$ are empty but site
$k_{3}$ is occupied, then the particle will move into site $k_{1}+1$
with probability $(1-q_{1})dt$, and:

\indent$\star$ If site $k_{2}-1$ is empty, the particle in site
$k_{1}$ will move into site $k_{2}$ with probability $q_{1}dt$.

\indent$\star$ If site $k_{2}-1$ is occupied, the particle in site
$k_{1}$ will move into site $k_{2}$ with probability $q_{1}p_{1}dt$.

$\bullet$ If site $k_{1}+1$ and site $k_{3}$ are empty but site
$k_{2}$ is occupied, then the particle will move into site $k_{1}+1$
with probability $(1-q_{2})dt$, and:

\indent$\star$ If site $k_{3}-1$ is empty, the particle in site
$k_{1}$ will move into site $k_{3}$ with probability $q_{2}dt$.

\indent$\star$ If site $k_{3}-1$ is occupied, the particle in site
$k_{1}$ will move into site $k_{3}$ with probability $q_{2}p_{2}dt$.

$\bullet$ If site $k_{2}$ and site $k_{3}$ are empty but site
$k_{1}+1$ is occupied, then:\\
\indent$\star$ If site $k_{2}-1$ is empty, the particle in site
$k_{1}$ will move into site $k_{2}$ with probability $q_{1}dt$.

\indent$\star$ If site $k_{2}-1$ is occupied, the particle in site
$k_{1}$ will move into site $k_{2}$ with probability $q_{1}p_{1}dt$.

\indent$\star$ If site $k_{3}-1$ is empty, the particle in site
$k_{1}$ will move into site $k_{3}$ with probability $q_{2}dt$.

\indent$\star$ If site $k_{3}-1$ is occupied, the particle in site
$k_{1}$ will move into site $k_{3}$ with probability $q_{2}p_{2}dt$.

$\bullet$ If sites $k_{1}+1$, $k_{2}$ and $k_{3}$ are all empty,
then the particle in site $k_{1}$ will move into site $k_{1}+1$ with
probability $(1-q_{1}-q_{2})dt$, and:

\indent$\star$ If site $k_{2}-1$ is empty, the particle in site
$k_{1}$ will move into site $k_{2}$ with probability $q_{1}dt$.

\indent$\star$ If site $k_{2}-1$ is occupied, the particle in site
$k_{1}$ will move into site $k_{2}$ with probability $q_{1}p_{1}dt$.

\indent$\star$ If site $k_{3}-1$ is empty, the particle in site
$k_{1}$ will move into site $k_{3}$ with probability $q_{2}dt$.

\indent$\star$ If site $k_{3}-1$ is occupied, the particle in site
$k_{1}$ will move into site $k_{3}$ with probability $q_{2}p_{2}dt$.

If site $k_{2}-1$ has a particle, and site $k_{2}$ is empty, then
the particle will choose the way
as following:

$\bullet$ if site $k_{1}$ is empty, then the particle in site
$k_{2}-1$
will move into site $k_{2}$ with probability $dt$.

$\bullet$ if site $k_{1}$ is occupied, then the particle in site
$k_{2}-1$
will move into site $k_{2}$ with probability $(1-p_{1}q_{1})dt$.

If site $k_{3}-1$ has a particle, and site $k_{3}$ is empty, then
the particle will choose the way
as following:

$\bullet$ if site $k_{1}$ is empty, then the particle in site
$k_{3}-1$
will move into site $k_{3}$ with probability $dt$.

$\bullet$ if site $k_{1}$ is occupied, then the particle in site
$k_{3}-1$
will move into site $k_{3}$ with probability $(1-p_{2}q_{2})dt$.

\section{Analysis of models}
\subsection{Theoretical Analysis of the Basic Model}
In this section, we will give a theoretical analysis of the phase situation of
three segments of our basic model. We choose situation 1 of our basic model
described above as the base of our analysis.

According to figure 1(b), we are able to express the insertion and
extraction rates using densities of certain sites. These rates are the following:
\begin{equation}
\begin{array}{l}
\alpha_{eff1}=\alpha\\
\beta_{eff1}=(1-\rho_{k_{1}+1})\rho_{k_{2}}+(1-\rho_{k_{2}})(1-\rho_{k_{1}+1})(1-q)+q(1-\rho_{k_{2}})[(1-\rho_{k_{2}-1})+p\rho_{k_{2}-1}]\\
\alpha_{eff2}=\rho_{k_{1}}[\rho_{k_{2}}+(1-\rho_{k_{2}})(1-q)]\\
\beta_{eff2}=(1-\rho_{k_{2}})[(1-\rho_{k_{1}})+q(1-p)\rho_{k_{1}}+(1-q)\rho_{k_{1}}]\\
\alpha_{eff3}=\rho_{k_{1}}(1-\rho_{k_{2}-1})q+\rho_{k_{2}-1}(1-\rho_{k_{1}})+\rho_{k_{1}}\rho_{k_{2}-1}\\
\beta_{eff3}=\beta\\
J_{sc}=q\rho_{k_{1}}(1-\rho_{k_{2}})[1-\rho_{k_{2}-1}+p\rho_{k_{2}-1}]
\end{array}
\end{equation}
Certainly, the flux should be conserved:
\begin{equation}
J_{1}=J_{2}+J_{sc}=J_{3}
\end{equation}
Firstly, we are going to prove that segment 1 and segment 3 should be
in the same phase.

If segment 1 is in maximum-current phase, then we have
$J_{1}=\frac{1}{4}$ according to (3). According to (5), we have
$J_{3}=\frac{1}{4}$. So segment 3 is also in maximum-current
phase.

Secondly, we claim that the following four cases are impossible
$$(L,L,H),(L,H,H),(H,L,L),(H,H,L)$$
To prove that, we firstly focus on the first two situations.
If the phase condition is $(L,*,H)$, due to (1), we have:
\begin{equation}
\rho_{1}=\alpha \quad\quad
\rho_{L}=1-\beta,\quad\quad(\alpha,\beta<0.5)
\end{equation}
Because $J_{1}=J_{3}$, we obtain:
\begin{equation}
\alpha(1-\alpha)=\beta(1-\beta)
\end{equation}
When $\alpha\neq\beta$, from (7) we have $\alpha+\beta=1$, which
contradicts $\alpha,\beta<0.5$. Thus the phase situations $(L,*,H)$ are impossible.

If the phase condition is $(H,L,L)$, it is difficult for us to
analysis the original model with parameter $p$. So we set $p=1$,
which means that the particle in site $k_{1}$ has priority to move
into
site $k_{2}$ when site $k_{2}-1$ is occupied.

Due to (2), we have:
\begin{equation}
\beta_{eff1}=1-\rho_{k_{1}}=1-\rho_{k_{1}+1}+q\rho_{k_{1}+1}(1-\rho_{k_{2}})
\end{equation}
from which we can obtain:
\begin{equation}
\rho_{k_{1}}=\rho_{k_{1}+1}(1-q(1-\rho_{k_{2}}))
\end{equation}
Let $1-q(1-\rho_{k_{2}})=A$, then (9) turns into:
\begin{equation}
\rho_{k_{1}}=A\rho_{k_{1}+1}
\end{equation}
Due to (1), we have:
\begin{equation}
\alpha_{eff2}=\rho_{k_{1}+1}=\rho_{k_{1}}[1-q(1-\rho_{k_{2}})]
\end{equation}
The equation above can be turned into:
\begin{equation}
\rho_{k_{1}+1}=A\rho_{k_{1}}
\end{equation}
From (10) and (12), we reach:
\begin{equation}
\rho_{k_{1}}=A^{2}\rho_{k_{1}}
\end{equation}
If $\rho_{k_{1}}=0$, then $J_{1}=\rho_{k_{1}}(1-\rho_{k_{1}})=0$
does not agree with reality. So $\rho_{k_{1}}$ could not be 0. Thus
$A=1$. For all $q\in[0,1]$,
\begin{equation}
\rho_{k_{2}}=1
\end{equation}
On the other hand, due to (5), we have:
\begin{equation}
\rho_{k_{1}}(1-\rho_{k_{1}})=\rho_{k_{2}}(1-\rho_{k_{2}})
\end{equation}
Because $\rho_{k_{1}}=1-\beta_{eff1}>0.5, \rho_{k_{2}}<0.5$, we know
that $\rho_{k_{1}}\neq\rho_{k_{2}}$. Together with (15), we have
$\rho_{k_{1}}+\rho_{k_{2}}=1$. Together with (14), we get
$\rho_{k_{1}}=0$.
It is impossible. Thus the phase situation can't be $(H,L,L)$.

If the phase condition is $(H,H,L)$, also we can set $p=1$ for our
convenience to prove. Because segment 2 is in high-density phase, we
have $\beta_{eff2}=1-\rho_{k_{2}-1}$. Because segment 3 is in
low-density phase, we have $\alpha_{eff3}=\rho_{k_{2}}$. According
to
(4), we have the cases:
\begin{equation}
\begin{cases}
1-\rho_{k_{2}-1}=1-\rho_{k_{2}}-q\rho_{k_{1}}(1-\rho_{k_{2}})\\
\rho_{k_{2}}=q\rho_{k_{1}}(1-\rho_{k_{2}-1})+\rho_{k_{2}-1}
\end{cases}
\end{equation}
Let $q\rho_{k_{1}}=B$, then the cases become:
\begin{equation}
\begin{cases}
\rho_{k_{2}-1}=\rho_{k_{2}}+B(1-\rho_{k_{2}})\\
\rho_{k_{2}}=\rho_{k_{2}-1}+B(1-\rho_{k_{2}-1})
\end{cases}
\end{equation}
From (17) we can obtain:
\begin{equation}
(2B-B^{2})\rho_{k_{2}}=(2B-B^{2})
\end{equation}
Because $J_{3}=\rho_{k_{2}}(1-\rho_{k_{2}})$, we know that
$\rho_{k_{2}}\neq1$, together with (18), we will have $B=0$, which
leads to $\rho_{k_{1}}=0$, which does not agree with reality. Thus $(H,H,L)$ has been excluded.

Now we have proved that segment 1 and segment 3 should be in the
same phase. In the following, we will discuss which phase may segment 2 be in.

If the three segments are in $(L,*,L)$, from the proof which has
demonstrated that $(H,H,L)$ is impossible, we know that $(L,H,L)$
can not be true either. Thus we can definitely say that for
$(L,*,L)$, only $(L,L,L)$ can be right.

If the three segments are in $(H,*,H)$, let us see whether $(H,L,H)$ is true.
If the three segments are in $(H,L,H)$, then we have:
\begin{equation}
\begin{cases}
\beta_{eff1}=1-\rho_{k_{1}}=1-\rho_{k_{1}+1}+q\rho_{k_{1}+1}(1-\rho_{k_{2}})\\
\alpha_{eff2}=\rho_{k_{1}}[1-q(1-\rho_{k_{2}})]=\rho_{k_{1}+1}
\end{cases}
\end{equation}
Let $1-q(1-\rho_{k_{2}})=A$, then from (19) we can get
$\rho_{k_{1}}=A\rho_{k_{1}}$, then $A=1$, then $\rho_{k_{2}}=1$.

Because of (5), we have:
\begin{equation}
\rho_{k_{1}}(1-\rho_{k_{1}})=\rho_{k_{1}+1}(1-\rho_{k_{1}+1})+q\rho_{k_{1}}(1-\rho_{k_{2}})=\beta(1-\beta)
\end{equation}
If we put $\rho_{k_{2}}=1$ into (20), we will get:
\begin{equation}
\rho_{k_{1}}(1-\rho_{k_{1}})=\rho_{k_{1}+1}(1-\rho_{k_{1}+1})
\end{equation}
Because $\rho_{k_{1}}\neq\rho_{k_{1}+1}$, we have
$\rho_{k_{1}}+\rho_{k_{1}+1}=1$. From $\rho_{k_{1}}=1-\beta$, we
know that $\rho_{k_{1}+1}=\beta$, then $J_{2}=\beta(1-\beta)=J_{3}$,
which does not agree to reality. So $(H,L,H)$ is impossible.

Thus we have excluded $(H,L,H)$. We now know that for $(H,*,H)$,
only $(H,H,H)$ is possible.

As for $(M,*,M)$, it is a tough task for us to analyze the situation
of segment 2 using equations. But under assistance of numerical
simulations, it will be possible for us to find which phase is true
for segment 2.

For situation 2 of model A, we can also write the insertion rates
and extraction rates at special sites:
\begin{equation}
\begin{array}{l}
\alpha_{eff1}=\alpha\\
\beta_{eff1}=(1-\rho_{k_{1}+1})+q\rho_{k_{1}+1}(1-\rho_{k_{2}})[(1-\rho_{k_{2}-1})+p\rho_{k_{2}-1}]\\
\alpha_{eff2}=\rho_{k_{1}}[\rho_{k_{2}}+(1-\rho_{k_{2}})(1-q((1-\rho_{k_{2}-1})+p\rho_{k_{2}-1}))]\\
\beta_{eff2}=(1-\rho_{k_{2}})[(1-\rho_{k_{1}})+q(1-p)\rho_{k_{1}}+(1-q)\rho_{k_{1}}]\\
\alpha_{eff3}=\rho_{k_{1}}(1-\rho_{k_{2}-1})q+\rho_{k_{2}-1}(1-\rho_{k_{1}})+\rho_{k_{1}}\rho_{k_{2}-1}\\
\beta_{eff3}=\beta\\
\end{array}
\end{equation}
The analysis of situation 2 of basic model is the same of the
analysis of situation 1, providing $p=1$, so we do not state it repeatedly.

From the analysis above, we can conclude that for the situation
$p=1$, the three segments should be among the following situations:
$(L,L,L),(H,H,H)$, $(M,*,M)$. We will do further investigation using
numerical simulations in the following sections.

\subsection{Analysis of advanced model 1}
In this section, we will give the phase situation of the
five segments of advanced model 1. According to figure 2(b), we are
able to express the insertion and extraction rates using densities
of certain sites. These rates are the following:
\begin{equation}
\begin{array}{l}
\alpha_{eff1}=\alpha\\
\beta_{eff1}=(1-\rho_{k_{1}+1})\rho_{k_{2}}+(1-\rho_{k_{2}})(1-\rho_{k_{1}+1})(1-q_{1})+q_{1}(1-\rho_{k_{2}})[(1-\rho_{k_{2}-1})+p_{1}\rho_{k_{2}-1}]\\
\alpha_{eff2}=\rho_{k_{1}}[\rho_{k_{2}}+(1-\rho_{k_{2}})(1-q_{1})]\\
\beta_{eff2}=(1-\rho_{k_{2}})[(1-\rho_{k_{1}})+q_{1}(1-p_{1})\rho_{k_{1}}+(1-q_{1})\rho_{k_{1}}]\\
\alpha_{eff3}=\rho_{k_{1}}(1-\rho_{k_{2}-1})q_{1}+\rho_{k_{2}-1}(1-\rho_{k_{1}})+\rho_{k_{1}}\rho_{k_{2}-1}\\
\beta_{eff3}=(1-\rho_{k_{3}+1})\rho_{k_{4}}+(1-\rho_{k_{4}})(1-\rho_{k_{3}+1})(1-q_{2})+q_{2}(1-\rho_{k_{4}})[(1-\rho_{k_{4}-1})+p_{2}\rho_{k_{4}-1}]\\
\alpha_{eff4}=\rho_{k_{3}}[\rho_{k_{4}}+(1-\rho_{k_{4}})(1-q_{2})]\\
\beta_{eff4}=(1-\rho_{k_{4}})[(1-\rho_{k_{4}})+q_{2}(1-p_{2})\rho_{k_{3}}+(1-q_{2})\rho_{k_{3}}]\\
\alpha_{eff5}=\rho_{k_{3}}(1-\rho_{k_{4}-1})q_{2}+\rho_{k_{4}-1}(1-\rho_{k_{3}})+\rho_{k_{3}}\rho_{k_{4}-1}\\
\beta_{eff5}=\beta\\
J_{sc_{1}}=q_{1}\rho_{k_{1}}(1-\rho_{k_{2}})[1-\rho_{k_{2}-1}+p_{1}\rho_{k_{2}-1}]\\
J_{sc_{2}}=q_{2}\rho_{k_{3}}(1-\rho_{k_{4}})[1-\rho_{k_{4}-1}+p_{2}\rho_{k_{4}-1}]\\
\end{array}
\end{equation}
Also, the flux should satisfies:
\begin{equation}
J_{1}=J_{2}+J_{sc_{1}}=J_{3}=J_{4}+J_{sc_{2}}=J_{5}
\end{equation}
In the following, we will prove that the segment 1,3 and 5 are in the same phase.
Conclusions obtained in the proof of basic model will be used in the following steps.
For the sake of simplicity, we always set $p_{1}=p_{2}=1$.

If one of the segments 1, 3 or 5 is in maximum-current phase, then
the other two will also be in maximum-current phase, since $J_{1}=J_{3}=J_{5}=\frac{1}{4}$.
Due to (6) and (7), we also can find that the situation is $(L,*,*,*,H)$ is impossible.

In the following, we discuss the situation $(H,*,*,*,L)$.
If segment 3 is in low-density phase, then segment 1,2 and 3 will
be in $(H,*,L)$. According to the proof of the basic
model, we know that this situation is impossible. Because we can
regard segment 1,2 and 3 as basic model, only by changing $\beta$ to be
$\beta_{eff3}$. This change does not influence the proof
since the rate expression of $\beta_{eff3}$ is independent of the
densities of sites in the first three segments.

Similarly, segment 3 can not be in high-density phase
since segment 3,4 and 5 will be in $(H,*,L)$. So now we have excluded $(L,*,*,*,H)$ and $(H,*,*,*,L)$,.

On the other hand, if the phase situation is $(L,*,*,*,L)$, then we know immediately that
segment 3 can not be in high-density phase. So segment 1,3 and 5
should all be in low-density phase. If the situation is $(H,*,*,*,H)$, then we know immediately that
segment 3 can not be in low-density phase. So segment 1,3 and 5
should all be in high-density phase.

Since we have proved that $(L,H,L)$ and $(H,L,H)$ does not exist in
basic model, so we can conclude that all of the five segments should
all be in low-density phase, providing segment 1 is in low-density
phase, and all be in high-density phase, when segment 1 is in
high-density phase.

As for the situation $(M,*,M,*,M)$, it is difficult for us to
analyze it theoretically. So we will use numerical
simulations to see whether segment 2 and 4 are in high or
low-density phase. The simulations will be presented in the following section.

\subsection{Analysis of advanced model 2}
In this section, we are going to demonstrate the phase situation of
four segments of advanced model 2. According to figure 3(b), we are
able to express the insertion and extraction rates using densities
of certain sites. These rates are stated as follows:
\begin{equation}
\begin{array}{l}
\alpha_{eff1}=\alpha\\
\beta_{eff1}=(1-\rho_{k_{1}+1})\rho_{k_{2}}\rho_{k_{3}}+(1-\rho_{k_{2}})\rho_{k_{1}+1}\rho_{k_{3}}q_{1}[(1-\rho_{k_{2}-1})+p_{1}\rho_{k_{2}-1}]\\
\indent\indent +(1-\rho_{k_{3}})\rho_{k_{1}+1}\rho_{k_{2}}q_{2}[(1-\rho_{k_{3}-1})+p_{2}\rho_{k_{3}-1}]\\
\indent\indent+(1-\rho_{k_{1}+1})(1-\rho_{k_{2}})\rho_{k_{3}}[1-q_{1}+q_{1}[(1-\rho_{k_{2}-1})+p_{1}\rho_{k_{2}-1}]]\\
\indent\indent+(1-\rho_{k_{1}+1})(1-\rho_{k_{3}})\rho_{k_{2}}[1-q_{2}+q_{2}[(1-\rho_{k_{3}-1})+p_{2}\rho_{k_{3}-1}]]\\
\indent\indent+\rho_{k_{1}+1}(1-\rho_{k_{2}})(1-\rho_{k_{3}})[q_{1}[(1-\rho_{k_{2}-1})+p_{1}\rho_{k_{2}-1}]+q_{2}[(1-\rho_{k_{3}-1})+p_{2}\rho_{k_{3}-1}]]\\
\indent\indent+(1-\rho_{k_{1}+1})(1-\rho_{k_{2}})(1-\rho_{k_{3}})[1-q_{1}-q_{2}+q_{1}[(1-\rho_{k_{2}-1})+p_{1}\rho_{k_{2}-1}]\\
\indent\indent+q_{2}[(1-\rho_{k_{3}-1})+p_{2}\rho_{k_{3}-1}]]\\
\alpha_{eff2}=\rho_{k_{1}}[\rho_{k_{2}}\rho_{k_{3}}+(1-\rho_{k_{2}})\rho_{k_{3}}(1-q_{1})+(1-\rho_{k_{3}})\rho_{k_{2}}(1-q_{2})\\
\indent\indent+(1-\rho_{k_{2}})(1-\rho_{k_{3}})(1-q_{1}-q_{2})]\\
\beta_{eff2}=(1-\rho_{k_{2}})[(1-\rho_{k_{1}})+q_{1}(1-p_{1})\rho_{k_{1}}+(1-q_{1})\rho_{k_{1}}]\\
\alpha_{eff3}=\rho_{k_{1}}(1-\rho_{k_{2}-1})q_{1}+\rho_{k_{2}-1}(1-\rho_{k_{1}})+\rho_{k_{1}}\rho_{k_{2}-1}\\
\beta_{eff3}=(1-\rho_{k_{3}})[(1-\rho_{k_{1}})+q_{2}(1-p_{2})\rho_{k_{1}}+(1-q_{2})\rho_{k_{1}}]\\
\alpha_{eff4}=\rho_{k_{1}}(1-\rho_{k_{3}-1})q_{2}+\rho_{k_{3}-1}(1-\rho_{k_{1}})+\rho_{k_{1}}\rho_{k_{3}-1}\\
\beta_{eff4}=\beta\\
J_{sc_{1}}=q_{1}\rho_{k_{1}}(1-\rho_{k_{2}})[1-\rho_{k_{2}-1}+p_{1}\rho_{k_{2}-1}]\\
J_{sc_{2}}=q_{2}\rho_{k_{1}}(1-\rho_{k_{3}})[1-\rho_{k_{3}-1}+p_{2}\rho_{k_{3}-1}]\\
\end{array}
\end{equation}
Also we have the equation of flux:
\begin{equation}
\begin{array}{c}
J_{1}=J_{4}=J_{3}+J_{sc_{2}}\\
J_{3}=J_{2}+J_{sc_{1}}\\
\end{array}
\end{equation}
Our conclusion is that the segments 1 and 4 are in the same phase.
In order to simplify our proof, we always set $p_{1}=p_{2}=1$.

Firstly, if segment 1 or segment 4 is in maximum-current phase, then
the other should also be in maximum-current phase due to (26).

If the phase condition is $(L,*,*,H)$, then due to
$\alpha(1-\alpha)=\beta(1-\beta)$, we know that when
$\alpha\neq\beta$ we have $\alpha+\beta=1$, which is contradictory to
$\alpha<0.5$ and $\beta<0.5$.

Now we discuss the situation $(H,*,*,L)$.
If the segment 3 is in low-density phase, then we have:
\begin{equation}
J_{1}=\rho_{k_{1}}(1-\rho_{k_{1}})=J_{3}+J_{sc_{2}}=\rho_{k_{3}}(1-\rho_{k_{3}})=J_{4}\\
\end{equation}
and
\begin{equation}
J_{3}=\rho_{k_{2}}(1-\rho_{k_{2}})
\end{equation}
If segment 2 is in low-density phase, then
\begin{equation}
\alpha_{eff2}=\rho_{k_{1}}=\rho_{k_{1}+1}(1-q_{1}(1-\rho_{k_{2}})-q_{2}(1-\rho_{k_{3}}))\\
\end{equation}
Let $1-q_{1}(1-\rho_{k_{2}})-q_{2}(1-\rho_{k_{3}})=A$, we
have:
\begin{equation}
\rho_{k_{1}}=A\rho_{k_{1}+1}
\end{equation}
Since segment 1 is in high-density phase, we have:
\begin{equation}
\beta_{eff1}=1-\rho_{k_{1}}=1-\rho_{k_{1}+1}+q_{1}\rho_{k_{1}+1}(1-\rho_{k_{2}})+q_{2}\rho_{k_{1}+1}(1-\rho_{k_{3}})
\end{equation}
So we have:
\begin{equation}
\rho_{k_{1}+1}=A\rho_{k_{1}}
\end{equation}
From (30) and (32), we know that:
\begin{equation}
\rho_{k_{1}}=A^{2}\rho_{k_{1}}
\end{equation}
From (33) we have $A=1$ because $\rho_{k_{1}}$ can not be 0. So
$\rho_{k_{2}}=\rho_{k_{3}}=1$ for all $q_{1},q_{2}\in(0,1]$, which
implies $J_{4}=0$.
So segment 2 can not be in low-density phase.
If segment 2 is in high-density phase, then we have:
\begin{equation}
\begin{array}{l}
\beta_{eff2}=1-\rho_{k_{2}-1}=1-\rho_{k_{2}}-q_{1}\rho_{k_{1}}(1-\rho_{k_{2}})\\
\alpha_{eff3}=\rho_{k_{2}}=q_{1}\rho_{k_{1}}(1-\rho_{k_{2}-1})+\rho_{k_{2}-1}
\end{array}
\end{equation}
From (34) we know that $\rho_{k_{1}}=0$ or $\rho_{k_{2}}=0$, which
will lead to $J_{1}=0$ or $J_{3}=0$ that does not make sense when
$q_{1}\neq0$. So we have excluded $(H,*,L,L)$.

If the segment 3 is in high-density phase, then
\begin{equation}
\begin{array}{l}
\beta_{eff3}=1-\rho_{k_{3}-1}=1-\rho_{k_{3}}-q_{2}\rho_{k_{1}}(1-\rho_{k_{3}})\\
\alpha_{ef4}=\rho_{k_{3}}=q_{2}\rho_{k_{1}}(1-\rho_{k_{3}-1})+\rho_{k_{3}-1}
\end{array}
\end{equation}
From (35) we know that $\rho_{k_{1}}=0$ or $\rho_{k_{3}}=0$, which
will lead to $J_{1}=0$ or $J_{4}=0$ that does not make sense.
$(H,*,H,L)$ is also impossible.

Now we have proved that segment 1 and segment 4 are in the same
phase. We now discuss the situation $(L,*,*,L)$. From the proof above we know
that segment 3 can not be in high-density phase, either can not
segment 2. Thus the phases can only be $(L,L,L,L)$.

While considering $(H,*,*,H)$, we should first assume that segment 3
be in low-density phase. From the proof above we will know that
segment 2 can not be either in low or in high-density phase. And if
we assume segment 3 to be in high-density phase, we know that
segment 2 can not be in low-density phase. Thus the phases can only
be $(H,H,H,H)$.

To the situation $(M,*,*,M)$, it is rather difficult for us to analyze it theoretically. So we also use numerical
simulations to get the results.

\section{Numerical Simulations of the different models}
\subsection{Simulations of basic models}
In this section, we use numerical methods to simulate basic models.
First we demonstrate the results of situation 1 and situation 2.

%Now we use numerical methods to simulate situation 1 of basic model.
According to analysis in last section, we set insertion rate
$\alpha$ and extraction rate $\beta$ to be 0.3 and 0.8 for
$(L,L,L)$, 0.8 and 0.3 for $(H,H,H)$, 0.8 and 0.8 for $(M,*,M)$.
The results of the simulations are plotted in figure 4.
\begin{figure}
  \begin{minipage}[t]{0.5\linewidth}
    \centering
    \includegraphics[width=2.5in]{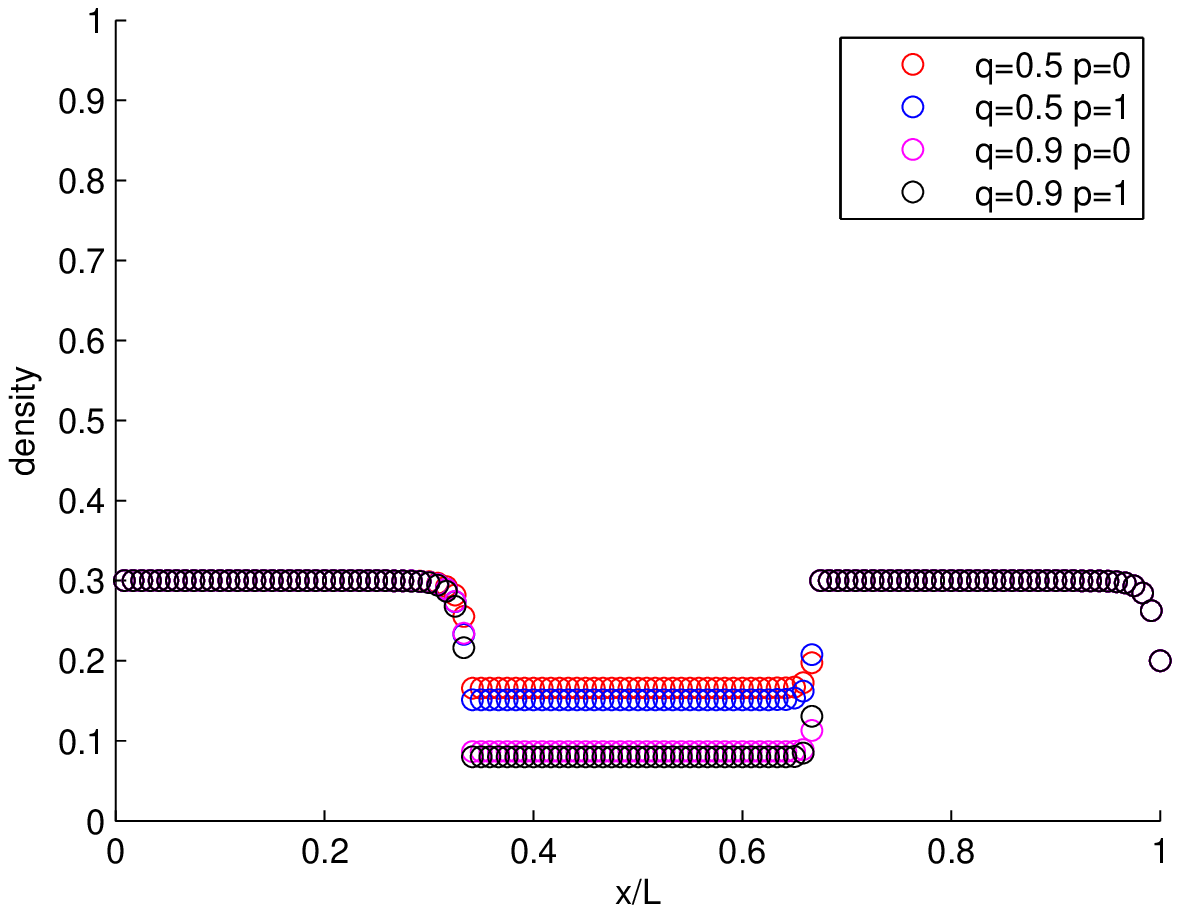}
    \label{fig:side:a}
    \subfigure{\footnotesize (a)}
  \end{minipage}
  \begin{minipage}[t]{0.5\linewidth}
    \centering
    \includegraphics[width=2.5in]{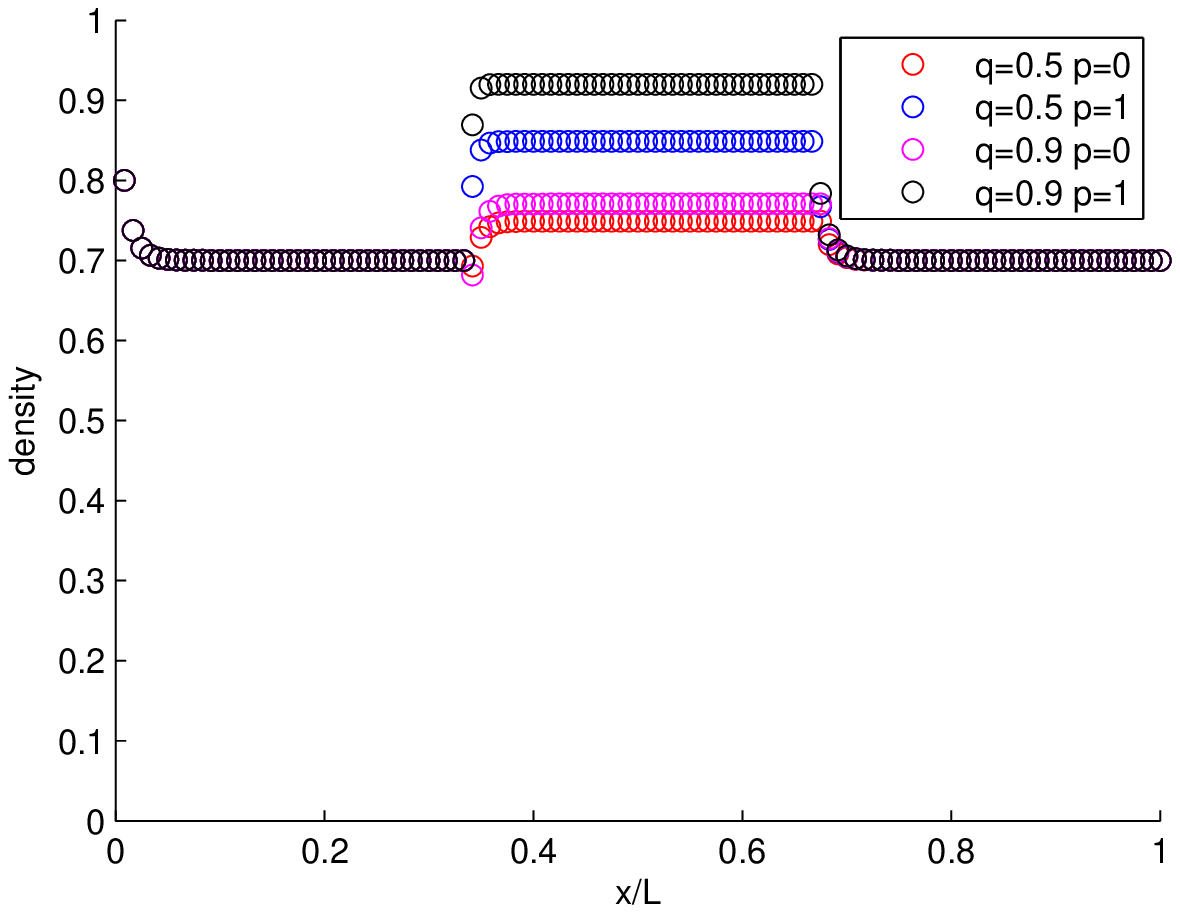}
    \label{fig:side:b}
    \subfigure{\footnotesize (b)}
  \end{minipage}
  \begin{minipage}[t]{0.5\linewidth}
    \centering
    \includegraphics[width=2.5in]{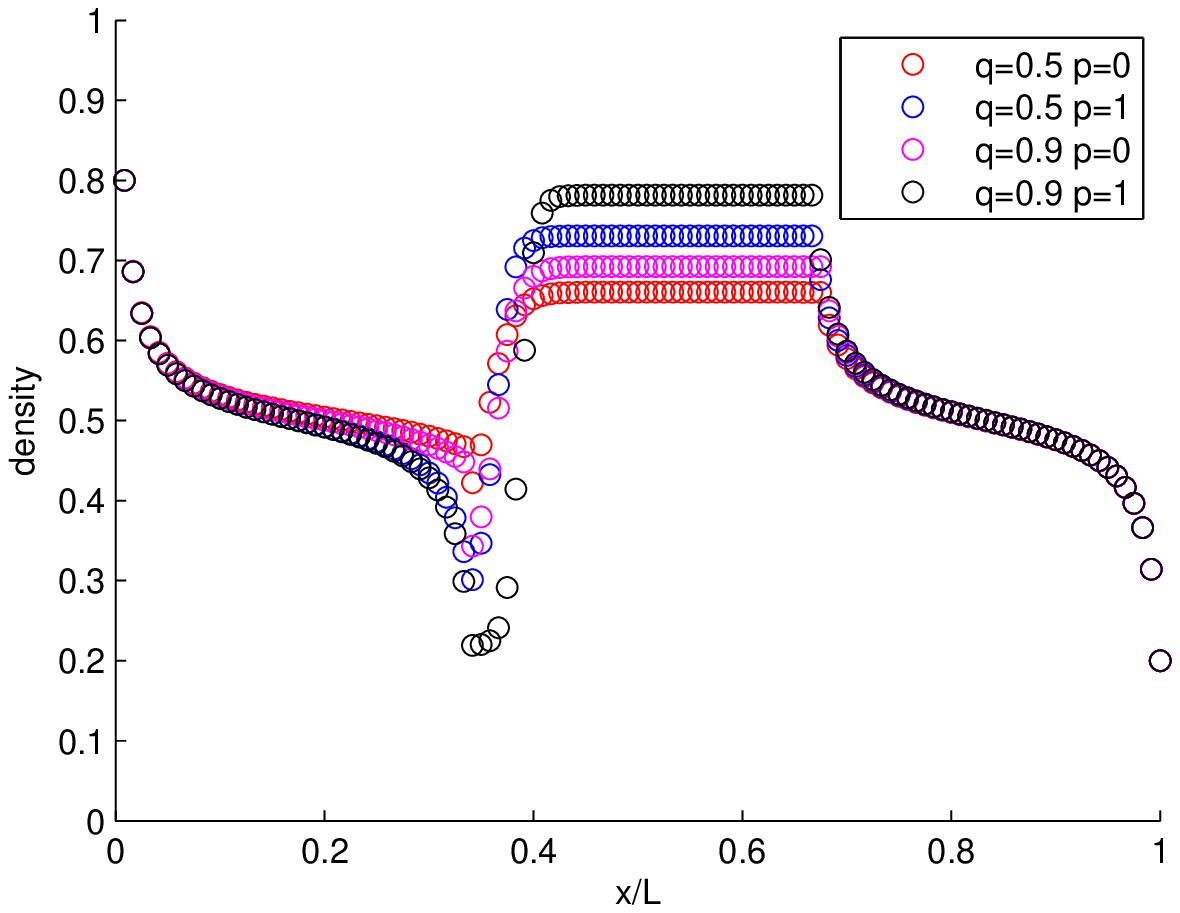}
    \label{fig:side:c}
    \subfigure{\footnotesize (c)}
  \end{minipage}
  \begin{minipage}[t]{0.5\linewidth}
    \centering
    \includegraphics[width=2.5in]{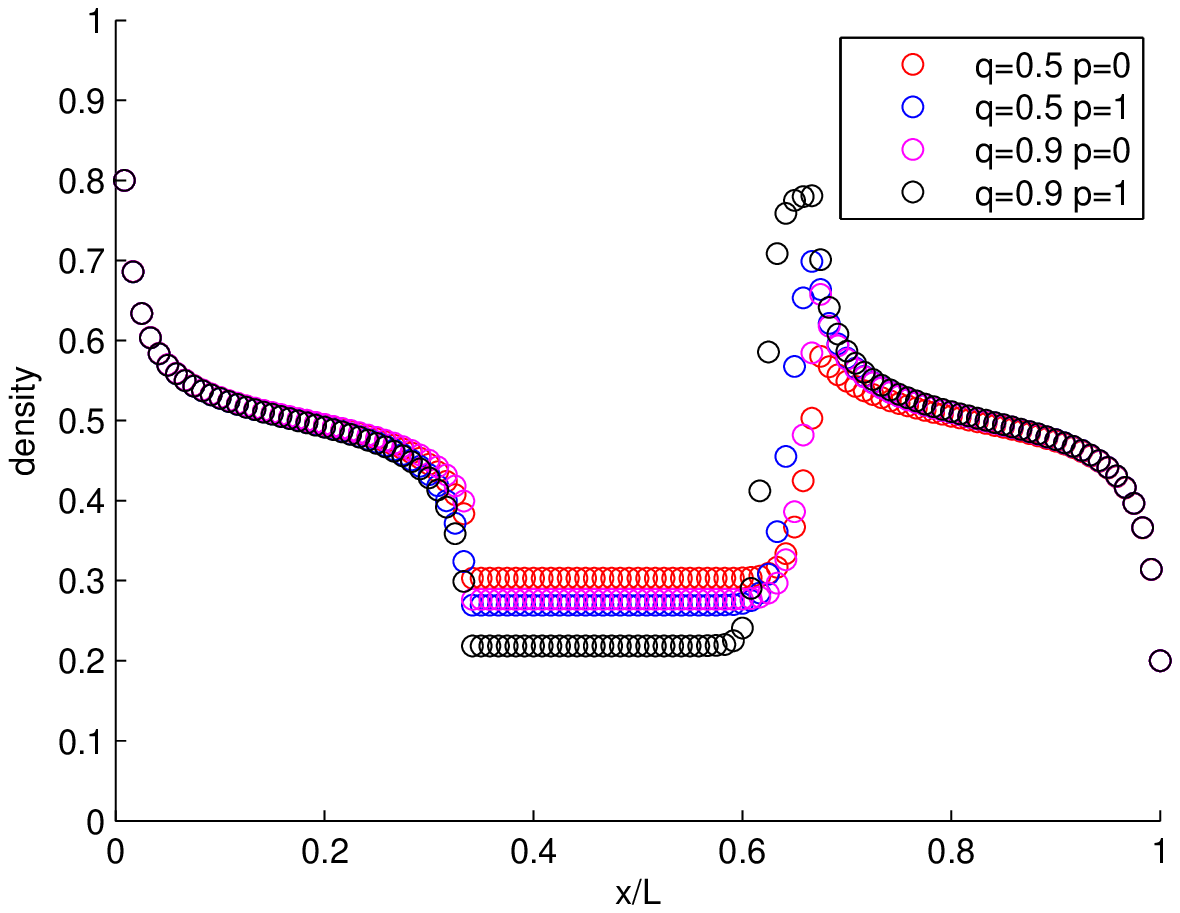}
    \label{fig:side:d}
    \subfigure{\footnotesize (d)}
  \end{minipage}
  \caption{\footnotesize Numerical simulations of basic model(situation 1) corresponding to different phases:(a)$\alpha=0.3,\beta=0.8$;
  (b)$\alpha=0.8,\beta=0.3$; (c)$\alpha=\beta=0.8$, initial densities of all sites are
  high; (d)$\alpha=\beta=0.8$, initial densities of all sites are low.}
\end{figure}

From figure 4, we can see that the results of numerical solutions
are in agreement with the results of equation analysis. When
$\alpha=0.3$ and $\beta=0.8$, all the three segments are in
low-density phase. When $\alpha=0.8$ and $\beta=0.3$, all the three
segments are in high-density phase. When $\alpha=\beta=0.8$, segment
1 and segment 3 are in maximum-current phase.

From figure 4(a), we can find that the density in segment 2 decreases
when $q$ increases, providing $p$ is constant. For the same $q$,
when $p$ increases, the density in segment 2 decreases.

From figure 4(b), we see that the density in segment 2 increases
when $q$ increases, providing $p$ is constant. For the same $q$,
when $p$ increases, the density in segment 2 increases. This is
because when $p$ is relatively large, more particles are able to
jump through the shortcut to reach site $k_{2}$, thus leaving
particles
in segment 2 in a serious traffic jam.

From figure 4(c) and (d), we see that when segment 1 and segment 3
are both in maximum-current phase, segment 2 can be either in
high-density phase or low-density phase, depending on the initial
density rate of each lattice. If the average initial density rate of
the lattices is high, then segment 2 will be in high-density phase
when the flux becomes stable. if the average initial density rate of
the lattices is low, then segment 2 will be in low-density phase
when the flux becomes stable.

Now we use numerical methods to simulate situation 2 of basic model.
The insertion and extraction rates are set as the same as
in the simulation of situation 1.
\begin{figure}
  \begin{minipage}[t]{0.5\linewidth}
    \centering
    \includegraphics[width=2.5in]{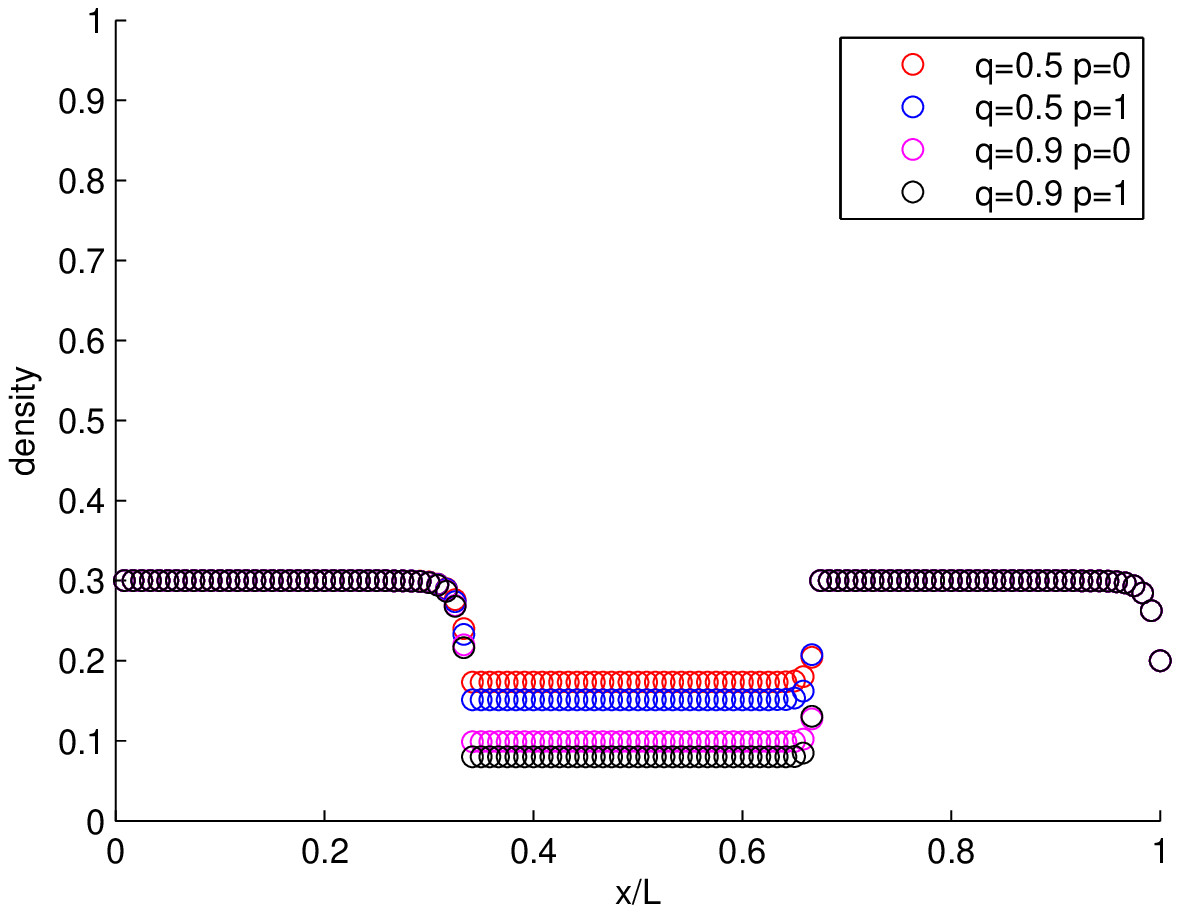}
    \label{fig:side:a}
    \subfigure{\footnotesize (a)}
  \end{minipage}
  \begin{minipage}[t]{0.5\linewidth}
    \centering
    \includegraphics[width=2.5in]{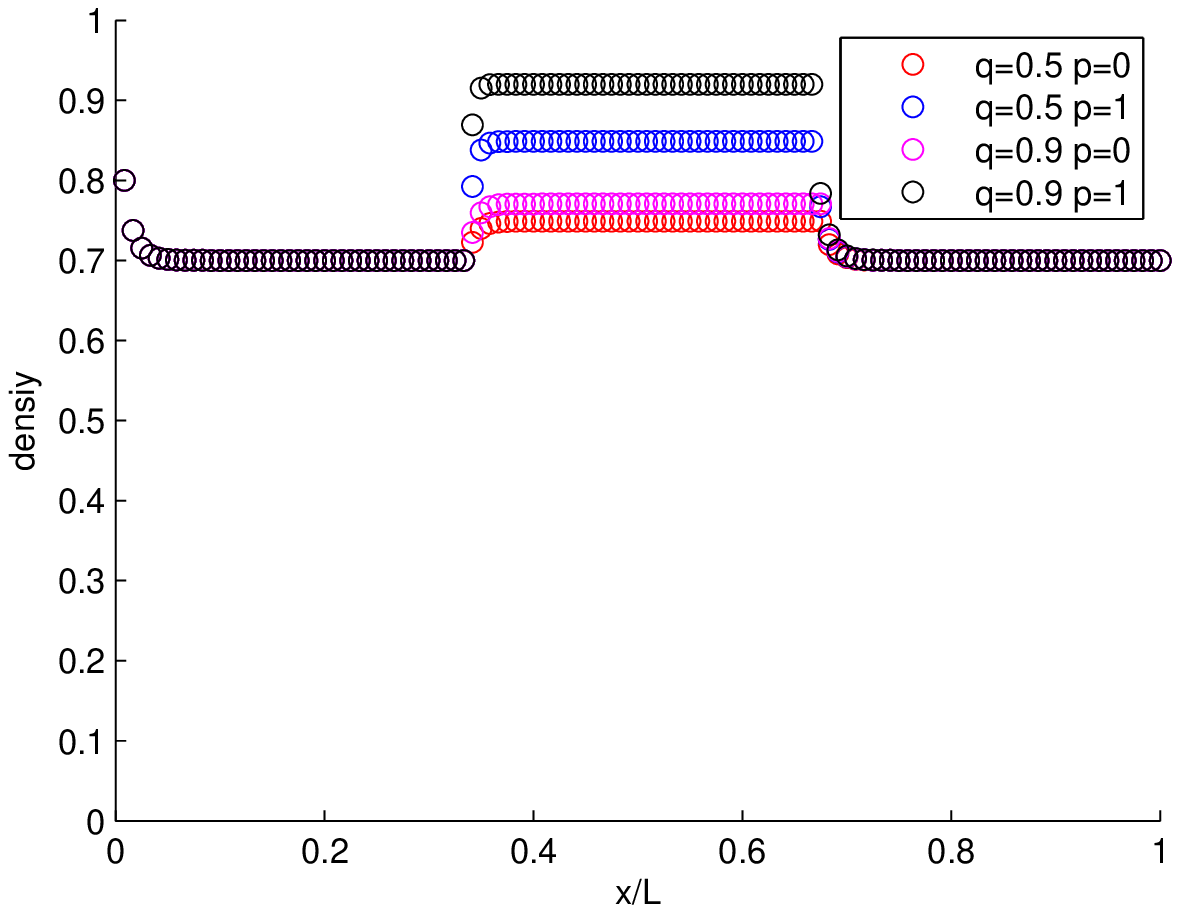}
    \label{fig:side:b}
    \subfigure{\footnotesize (b)}
  \end{minipage}
  \begin{minipage}[t]{0.5\linewidth}
    \centering
    \includegraphics[width=2.5in]{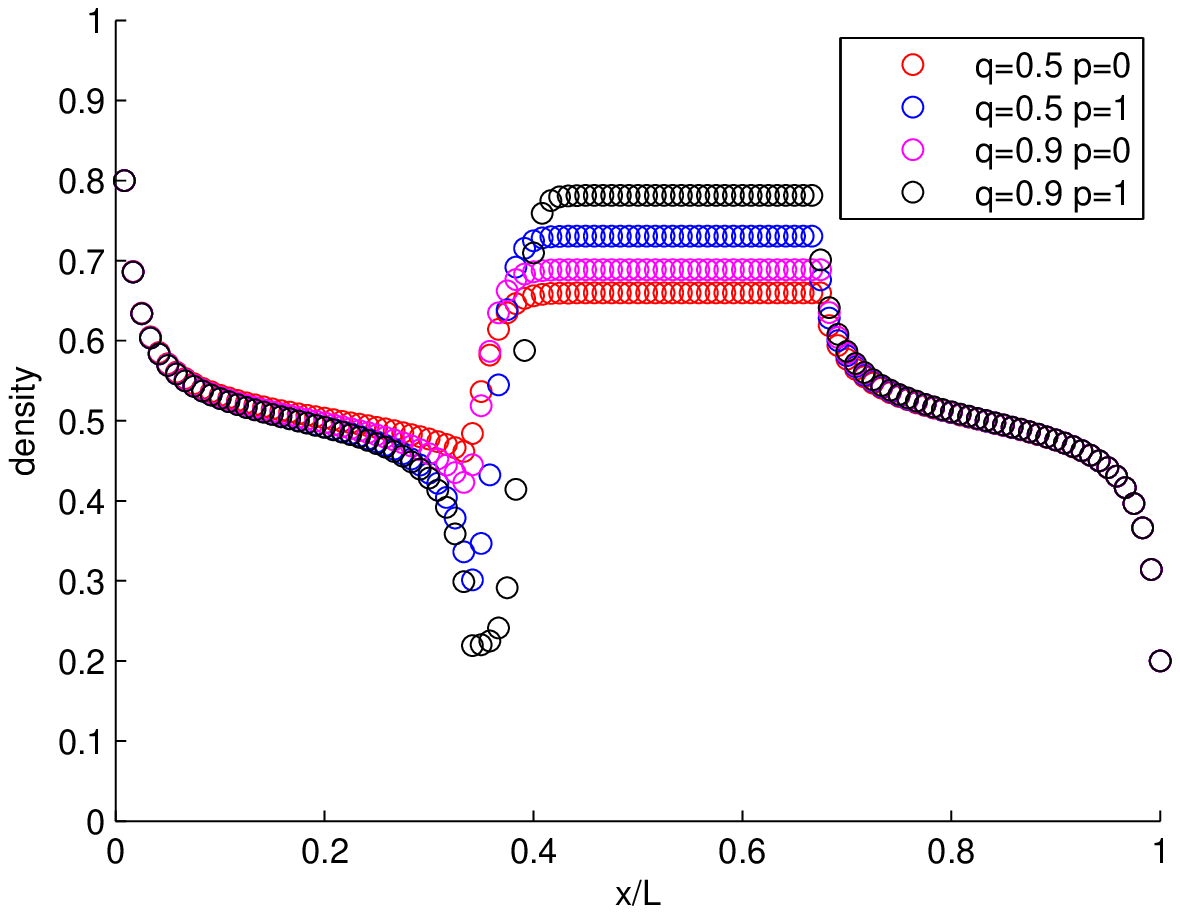}
    \label{fig:side:c}
    \subfigure{\footnotesize (c)}
  \end{minipage}
  \begin{minipage}[t]{0.5\linewidth}
    \centering
    \includegraphics[width=2.5in]{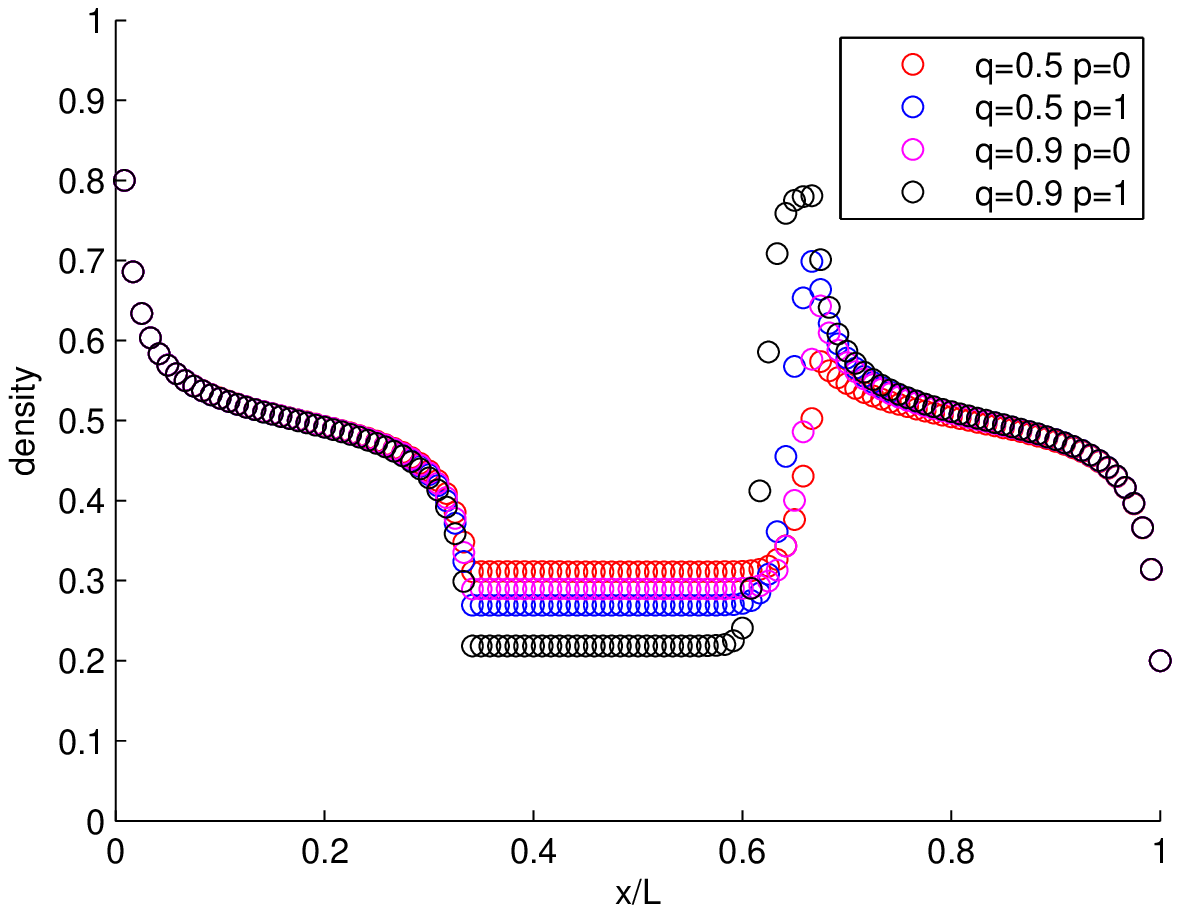}
    \label{fig:side:d}
    \subfigure{\footnotesize (d)}
  \end{minipage}
  \caption{\footnotesize numerical simulations of basic model(situation 2) corresponding to different phases:(a)$\alpha=0.3,\beta=0.8$;
  (b)$\alpha=0.8,\beta=0.3$; (c)$\alpha=\beta=0.8$, initial densities of all sites are
  high; (d)$\alpha=\beta=0.8$, initial densities of all sites are low.}
\end{figure}

From figure 5 we are able to see that the results of situation 2 of
basic model seem to be similar to the results of situation 1. But
after careful inspection, we may find that there are several
differences between the results of the two situations, which
manifest the distinct characters of these two situations.

In order to do further investigation, we use figures which can show
the differences between two situations. We calculate the values
which are obtained by subtracting the density of each lattice of
situation 2 from the density of each lattice of situation 1. When we
do these subtractions, we keep parameters $q$ and $p$ in constant.
The results are shown in figure 6.
\begin{figure}
  \begin{minipage}[t]{0.5\linewidth}
    \centering
    \includegraphics[width=2.5in]{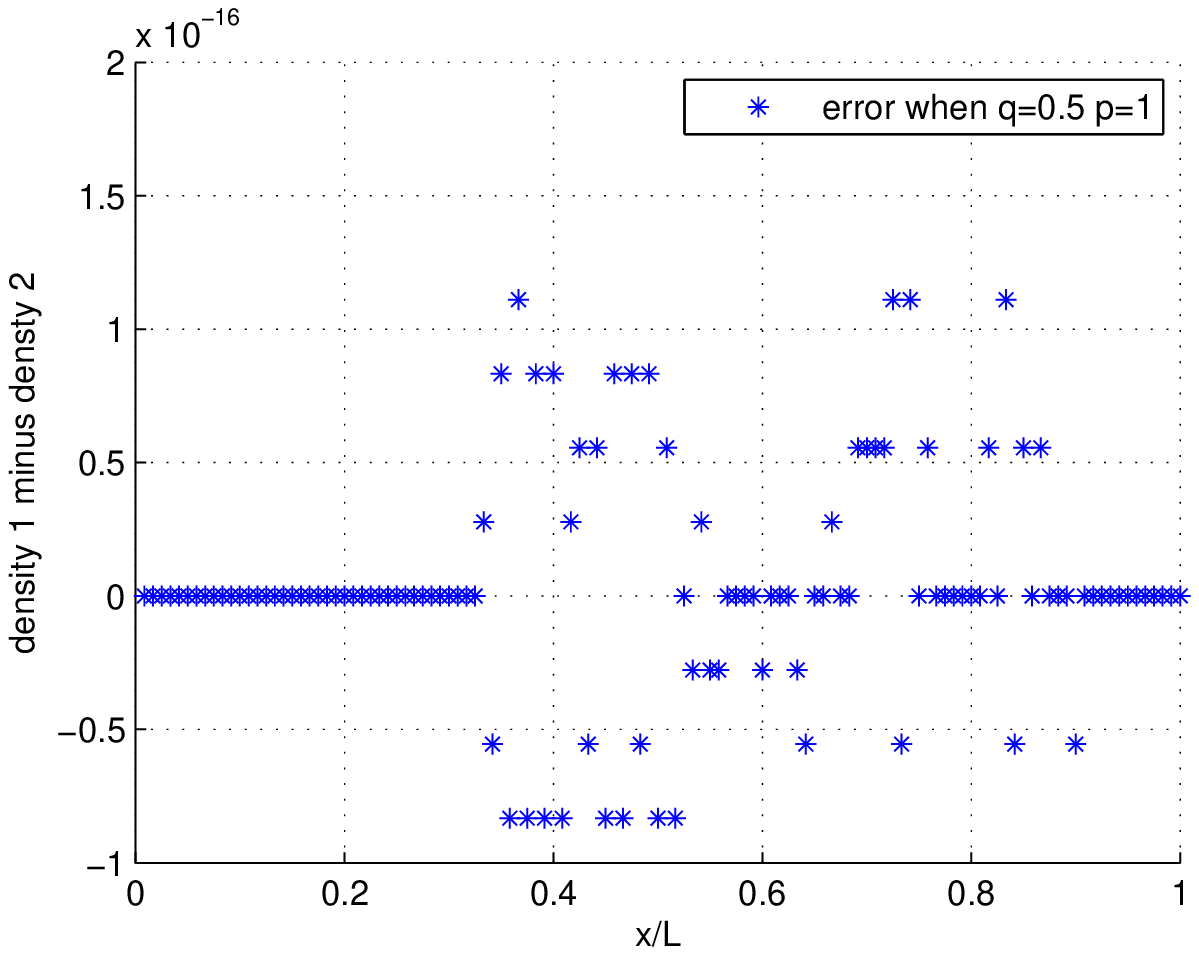}
    \label{fig:side:a}
    \subfigure{\footnotesize (a)}
  \end{minipage}
  \begin{minipage}[t]{0.5\linewidth}
    \centering
    \includegraphics[width=2.5in]{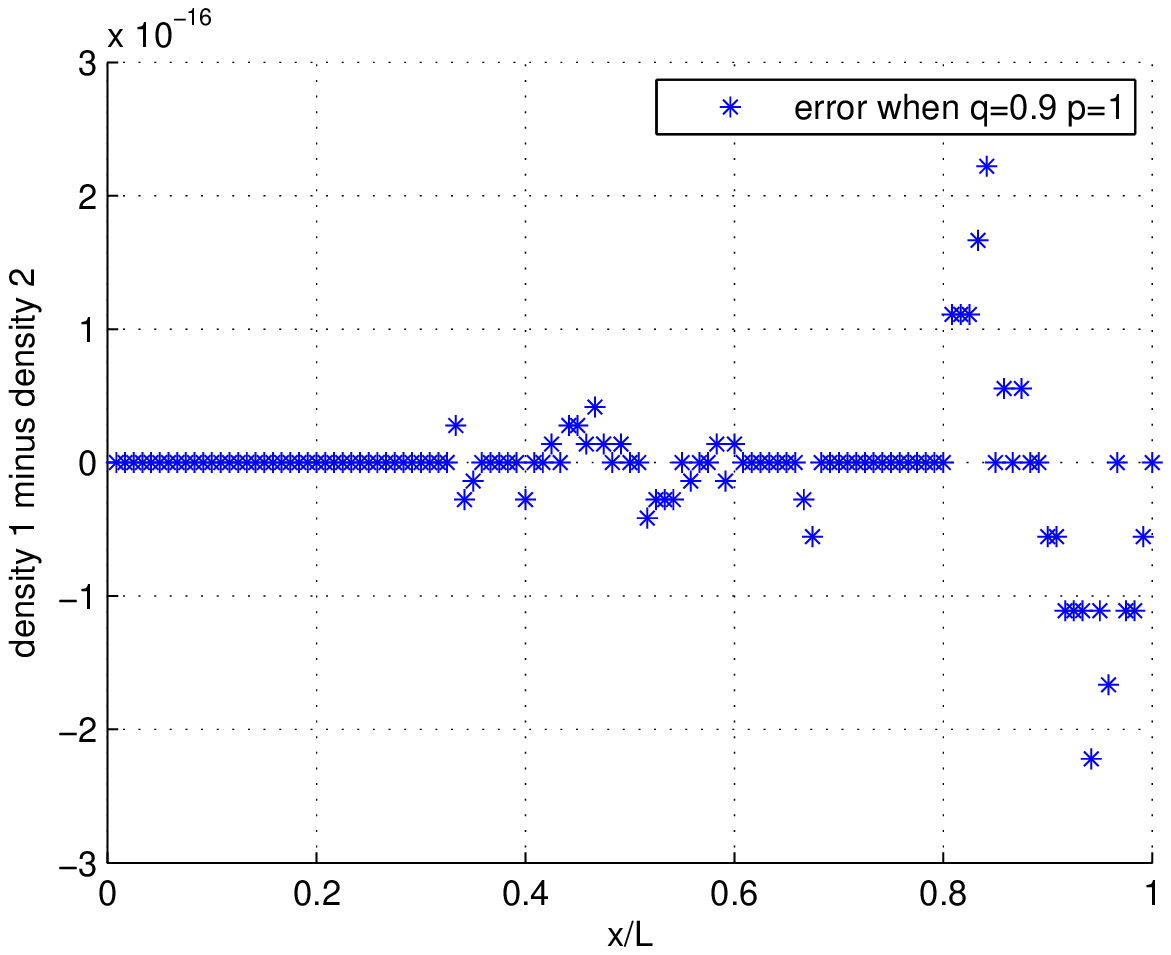}
    \label{fig:side:a}
    \subfigure{\footnotesize (b)}
  \end{minipage}
  \begin{minipage}[t]{0.5\linewidth}
    \centering
    \includegraphics[width=2.5in]{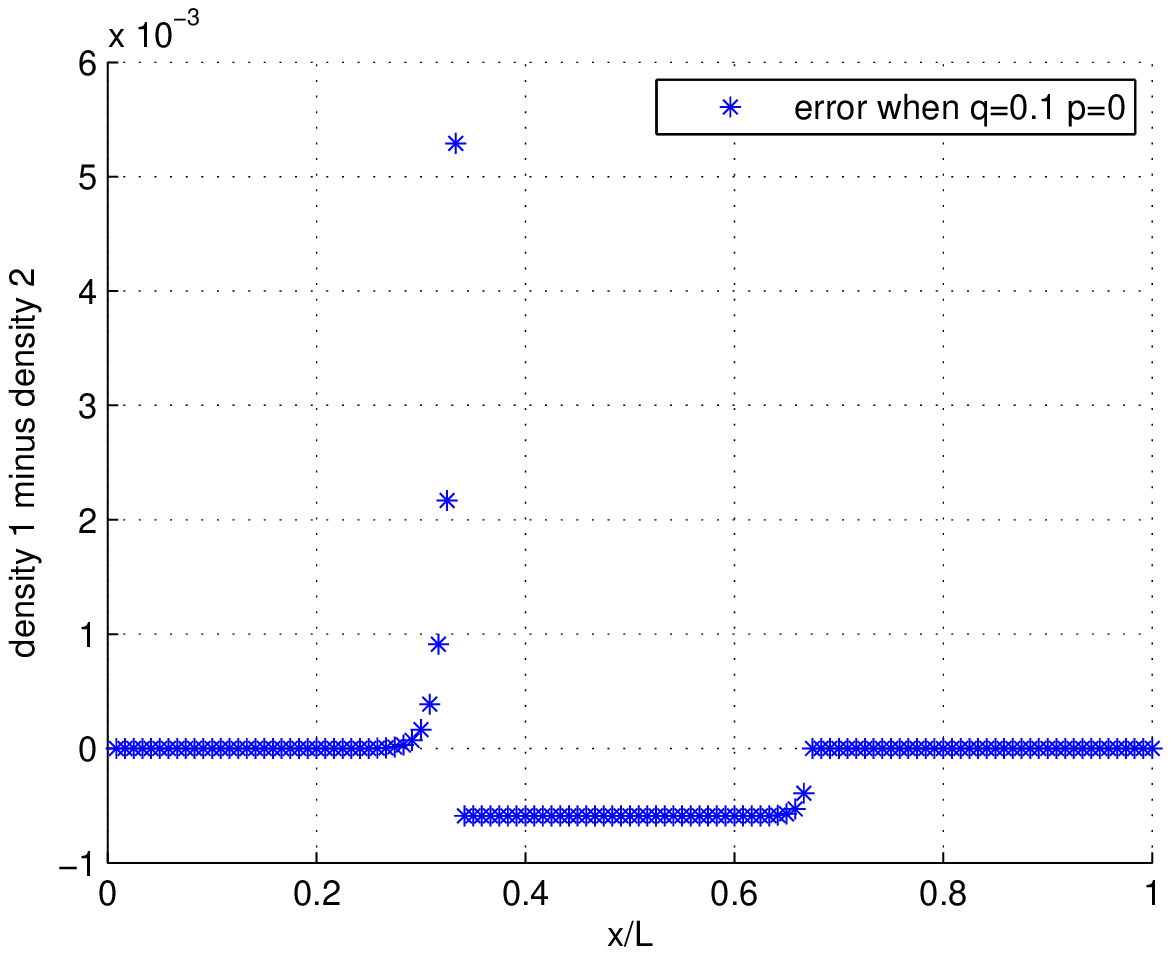}
    \label{fig:side:b}
    \subfigure{\footnotesize (c)}
  \end{minipage}
  \begin{minipage}[t]{0.5\linewidth}
    \centering
    \includegraphics[width=2.5in]{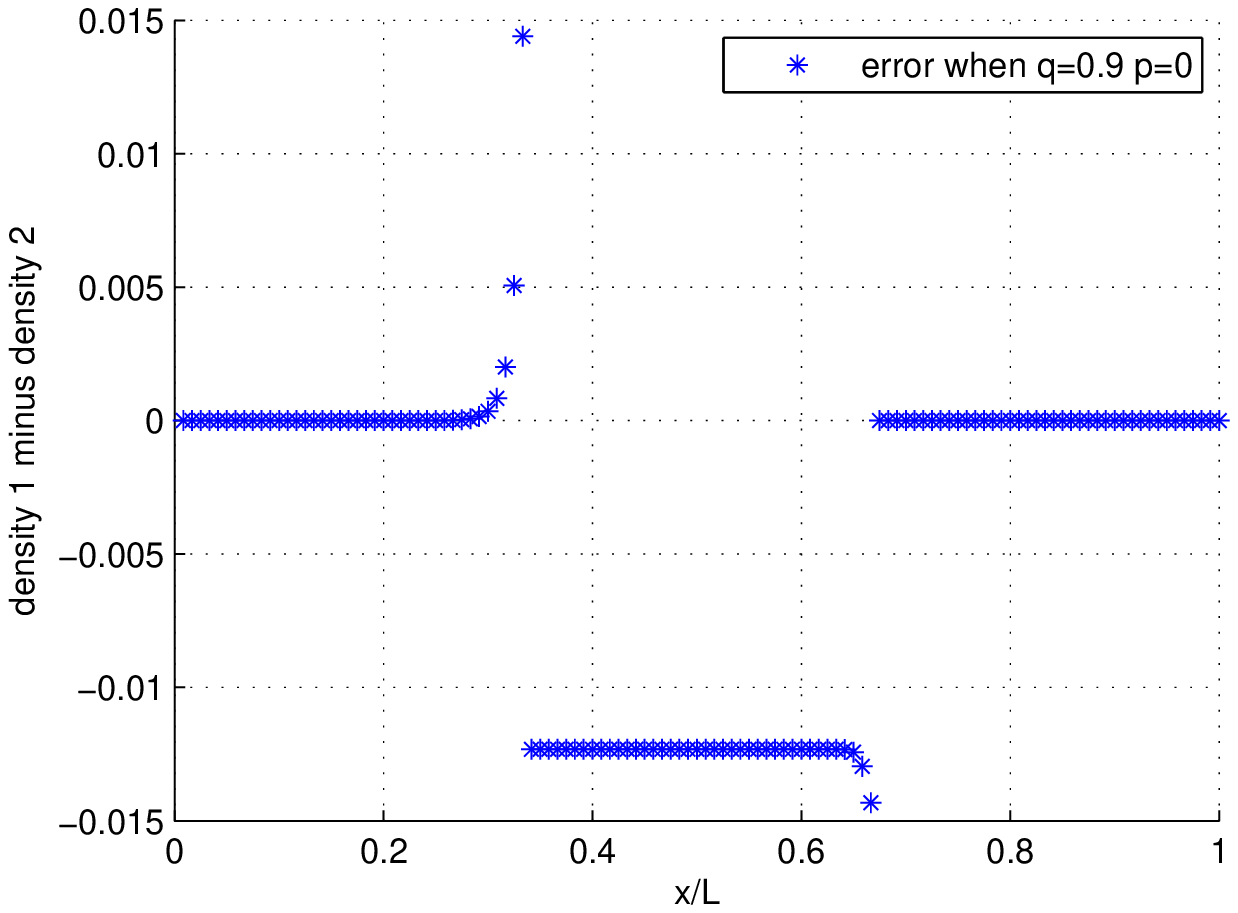}
    \label{fig:side:c}
    \subfigure{\footnotesize (d)}
  \end{minipage}
  \caption{\footnotesize The density of each lattice of situation 1 of basic model minus the density
   of each lattice of situation 2 of basic model with insertion rate $\alpha=0.3$ and extraction rate $\beta=0.8$:
   (a) when $q=0.5$ and $p=1$; (b) when $q=0.9$ and $p=1$; (c) when $q=0.1$ and $p=0$; (d) when $q=0.9$ and $p=0$.}
\end{figure}
In figure 6(a) and 6(b), we can see that the differences between the
density of two situations are within $3\times10^{-16}$, which means
that when $p=1$, there is nearly no difference between the two
situations. That is because when $p=1$, the particle at site
$k_{1}(x=\frac{1}{3}$ in the figure) has priority to get
to site $k_{2}(x=\frac{2}{3}$ in the  figure). So whatever
$q$ may be, the particle can go smoothly once it chooses the
shortcut. There is no
difference whether it returns back to segment 2 or not.

In figure 6(c) and (d), when $p=0$, the differences seem to be
obvious. when we contrast these two figures, we are able to find
that the extent of difference is decided by parameter $q$. While $q$
changes from 0.1 to 0.9, the error changes from within
$6\times10^{-3}$ to 0.015. This is simply because the smaller $q$
means less particles to choose the shortcut. But if we compare
figure 6(b) with 6(d), it can apparently be found that the error
changes greatly when $p$ is switched from 1 to 0, providing $q$ is
comparatively large. So we can say that it really matters whether a
particle can return back or not when many particles choose the
shortcut. If the particle can not return, a great traffic jam will
happen at the certain site, reflected in figure 6(d) as a
high-positioned blue star on the top of the figure.

\subsection{Simulations of advanced model 1}
Now we use numerical methods to simulate advanced model 1. In order
to find how the phases of five segments change when parameters
$q_{1}$ and $q_{2}$ change, we set $p_{1}=p_{2}=1$, thus
corresponding to our equation analysis in section 3. We will see
whether the numerical results correspond to our analytical
results.
\begin{figure}
  \begin{minipage}[t]{0.5\linewidth}
    \centering
    \includegraphics[width=2.5in]{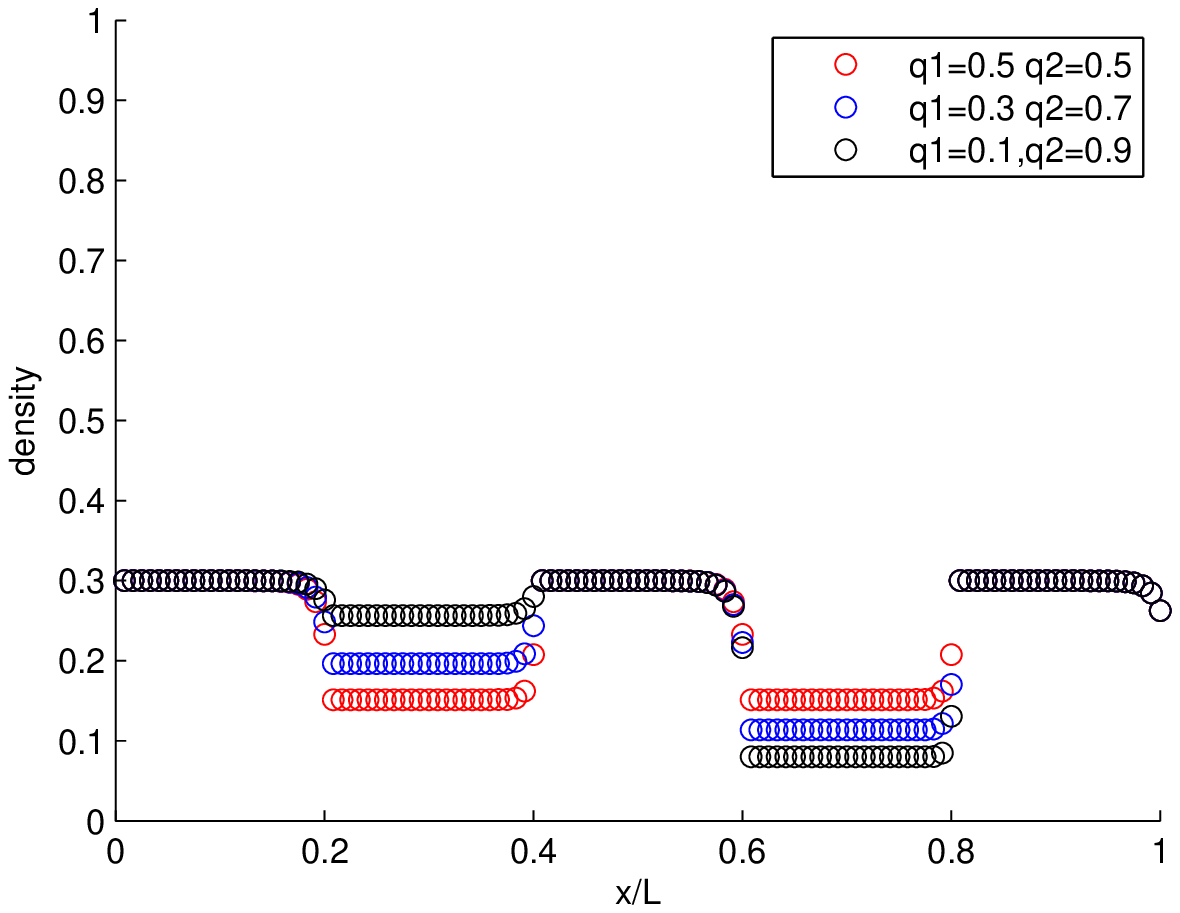}
    \label{fig:side:a}
    \subfigure{\footnotesize (a)}
  \end{minipage}
  \begin{minipage}[t]{0.5\linewidth}
    \centering
    \includegraphics[width=2.5in]{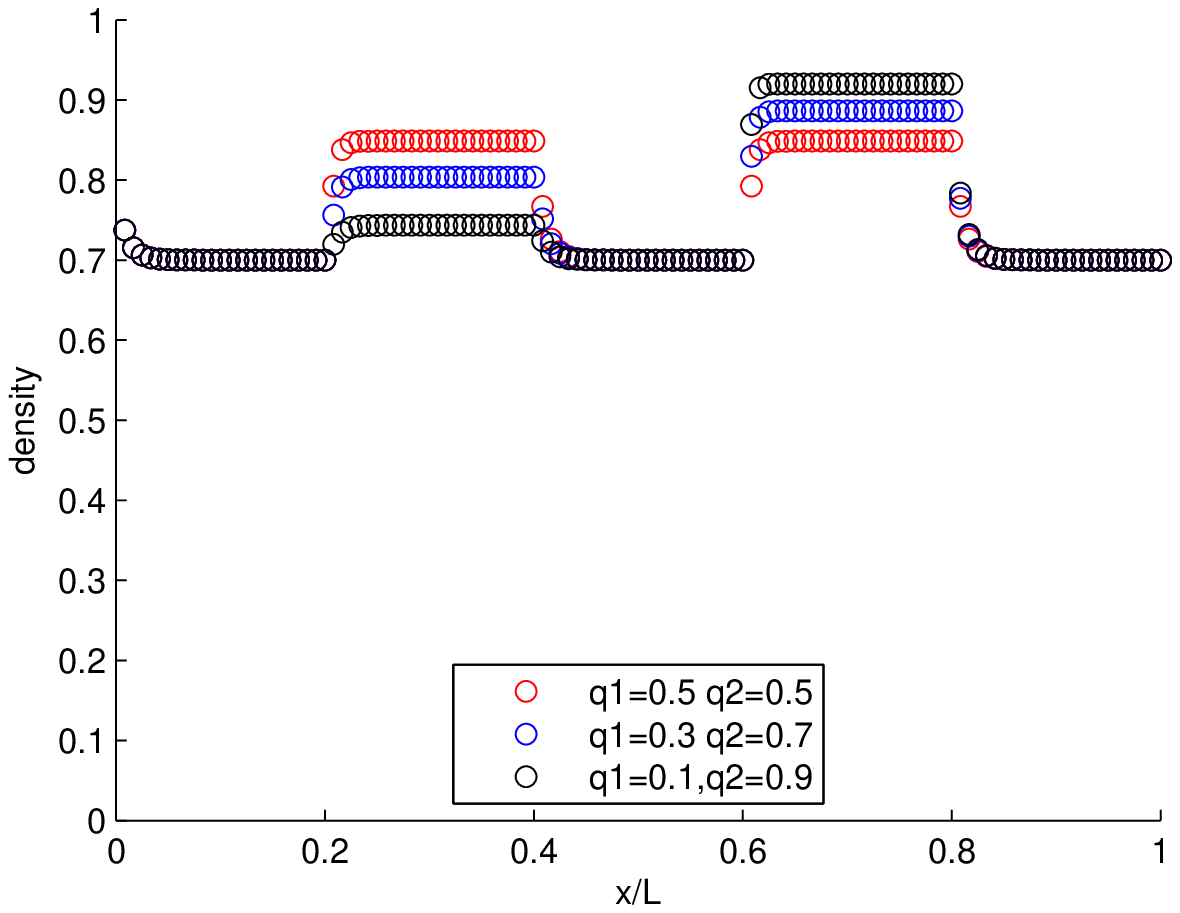}
    \label{fig:side:b}
    \subfigure{\footnotesize (b)}
  \end{minipage}
  \begin{minipage}[t]{0.5\linewidth}
    \centering
    \includegraphics[width=2.5in]{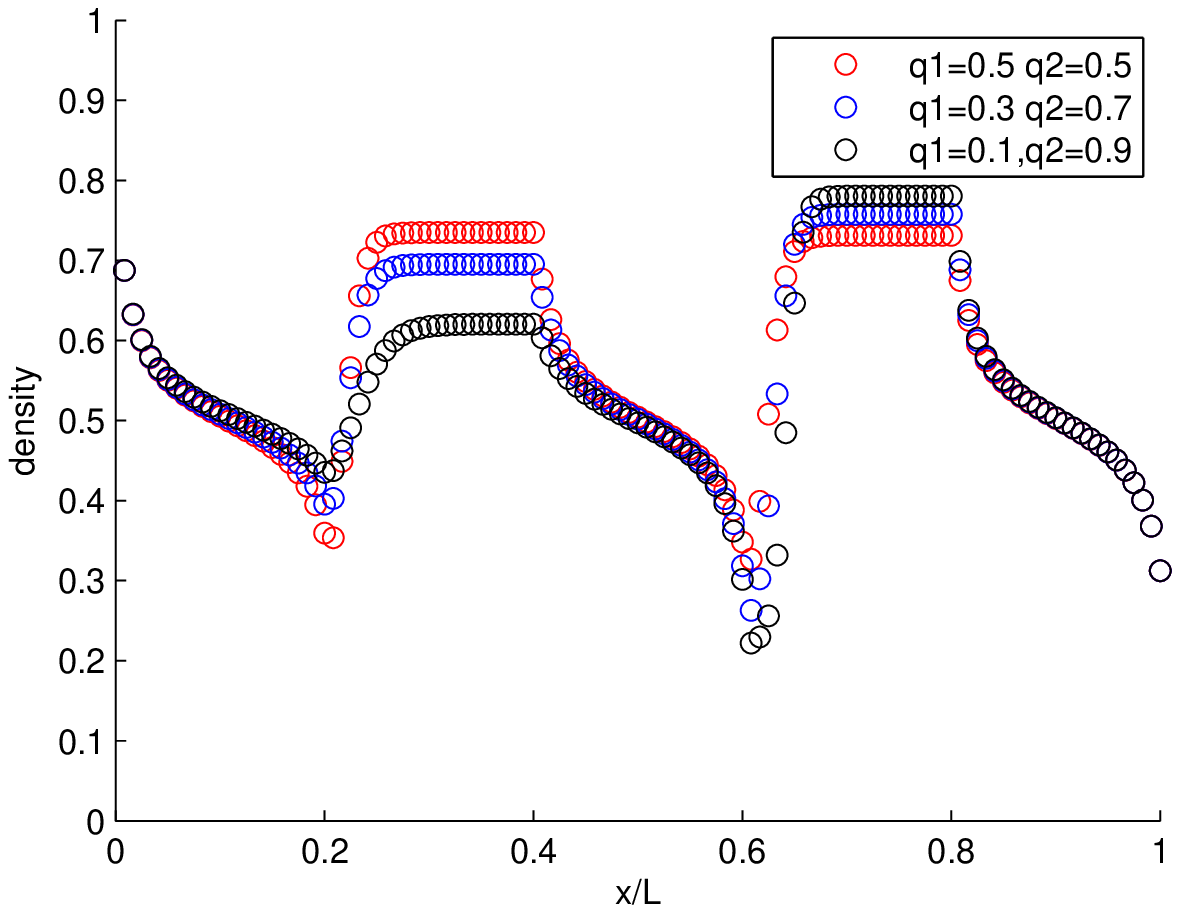}
    \label{fig:side:c}
    \subfigure{\footnotesize (c)}
  \end{minipage}
  \begin{minipage}[t]{0.5\linewidth}
    \centering
    \includegraphics[width=2.5in]{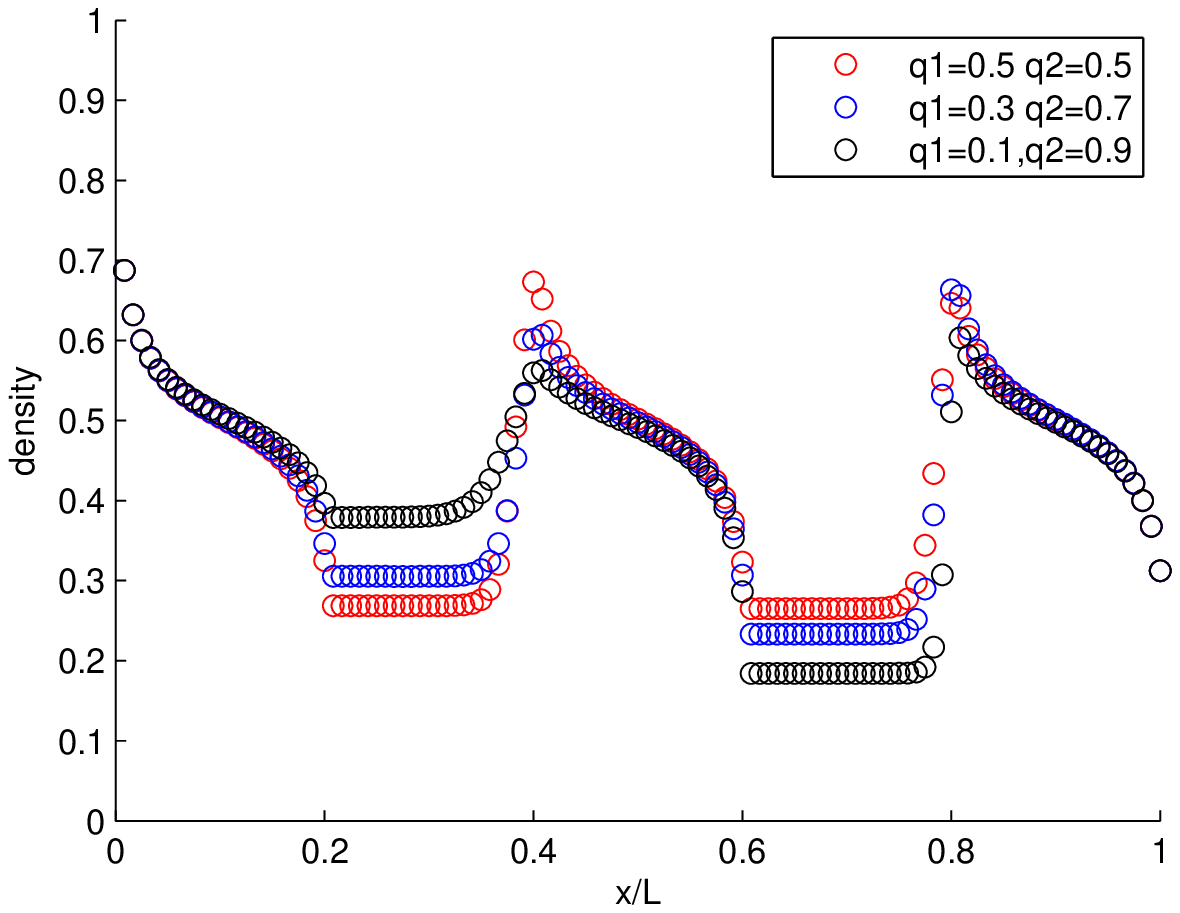}
    \label{fig:side:d}
    \subfigure{\footnotesize (d)}
  \end{minipage}
  \caption{\footnotesize numerical simulations of advanced model 1 corresponding to different phases:(a)$\alpha=0.3,\beta=0.8$;
  (b)$\alpha=0.8,\beta=0.3$; (c)$\alpha=\beta=0.8$, initial densities of all sites are
  high; (d)$\alpha=\beta=0.8$, initial densities of all sites are low.}
\end{figure}

From figure 7(a), we can see that all of the five segments are in
low-density phase. The situation in segment 2 has nothing to do with
the situation in segment 4. We can treat them as a connection of two
basic models. The density rate in segment 2 decreases when $q_{1}$
increases. The density rate in segment 4 decreases when $q_{2}$
increases.

From figure 7(b), we can see that all of the five segments are in
high-density phase. The situation in segment 2 has nothing to do
with the situation in segment 4. We can treat them as a connection
of two basic models. The density rate in segment 2 increases when
$q_{1}$ increases. The density rate in segment 4 increases when
$q_{2}$ increases.

From figure 7(c) and 7(d), we can see that the situation in segment
2 and 4 are relevant to the initial density of each site. If the
average initial density of all sites is high, then segment 2 and 4
will be both in high-density phase. If the average initial density
of all sites is low, then segment 2 and 4 will be both in
low-density
phase.

\subsection{Simulation of advanced model 2}
Now we use numerical methods to simulate advanced model 2. In order
to find how the phases of four segments change when parameters
$q_{1}$ and $q_{2}$ change, we set $p_{1}=p_{2}=1$, thus
corresponding to our equation analysis in chapter 3. We will see
whether the numerical results correspond to our analytical results.
Also, we are interested in whether there are interactions between
the two parameters $q_{1}$ and $q_{2}$.
\begin{figure}
  \begin{minipage}[t]{0.5\linewidth}
    \centering
    \includegraphics[width=2.5in]{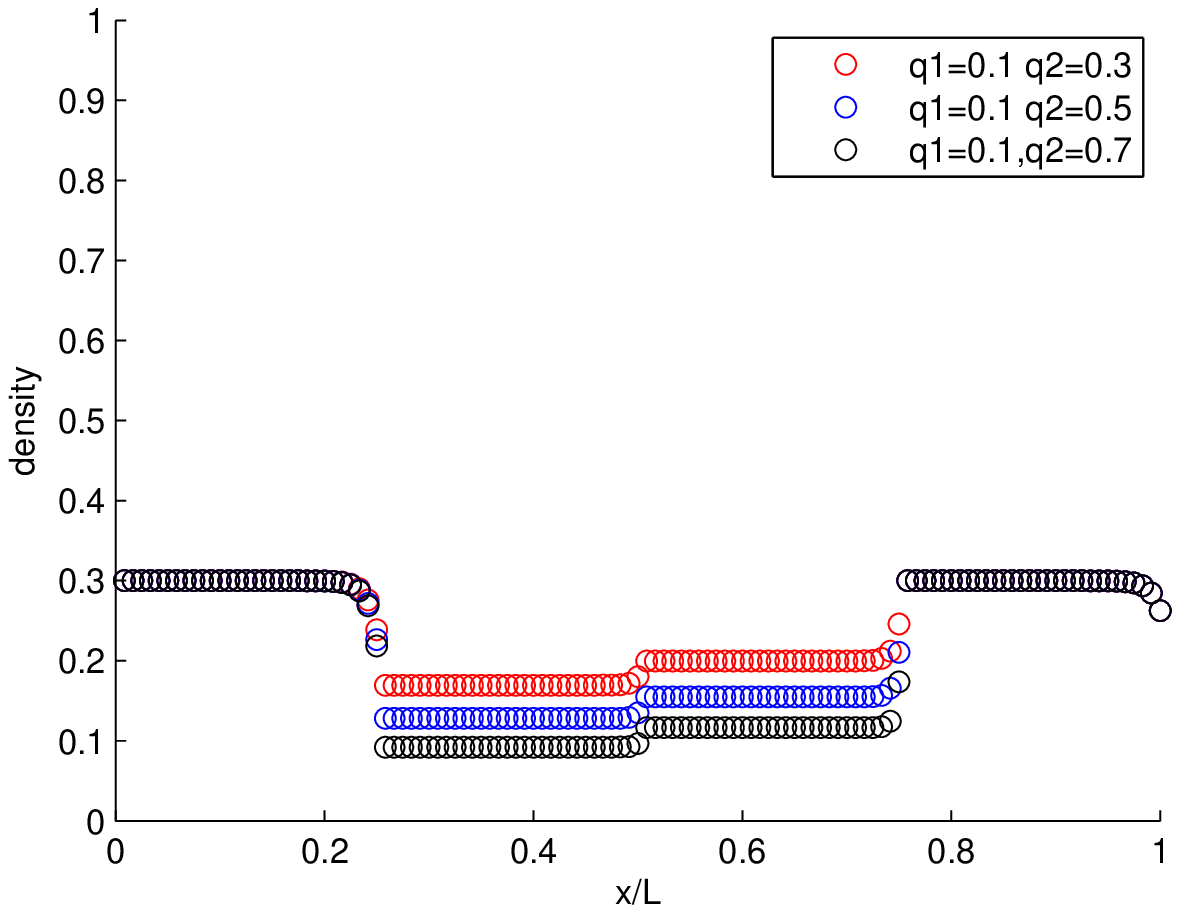}
    \label{fig:side:a}
    \subfigure{\footnotesize (a)}
  \end{minipage}
  \begin{minipage}[t]{0.5\linewidth}
    \centering
    \includegraphics[width=2.5in]{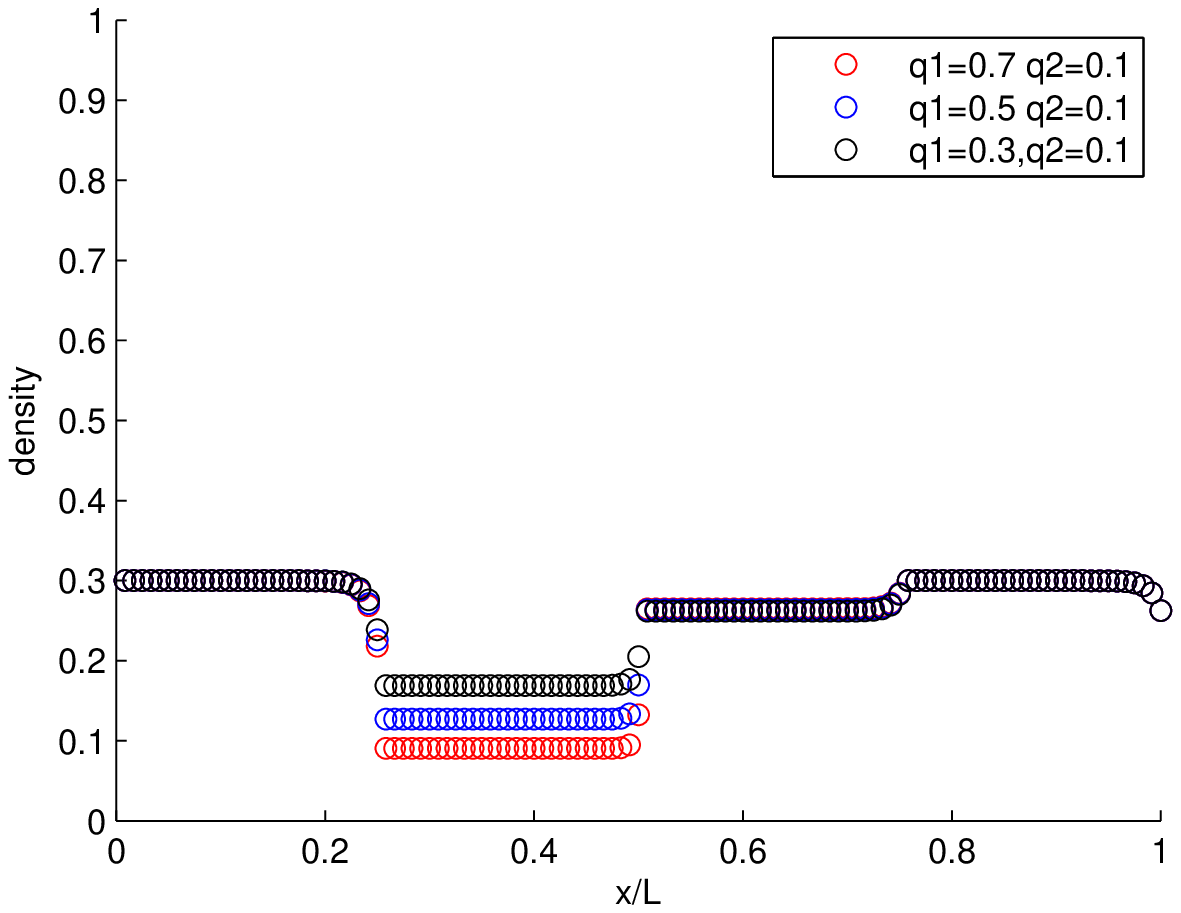}
    \label{fig:side:b}
    \subfigure{\footnotesize (b)}
  \end{minipage}
  \caption{\footnotesize numerical simulations of advanced model 2 with insertion rate $\alpha=0.3$ and
            $\beta=0.8$: (a)$q_{2}$ changes when $q_{1}$ remains constant; (b) $q_{1}$ changes when $q_{2}$ remains constant.}
\end{figure}
\begin{figure}
  \begin{minipage}[t]{0.5\linewidth}
    \centering
    \includegraphics[width=2.5in]{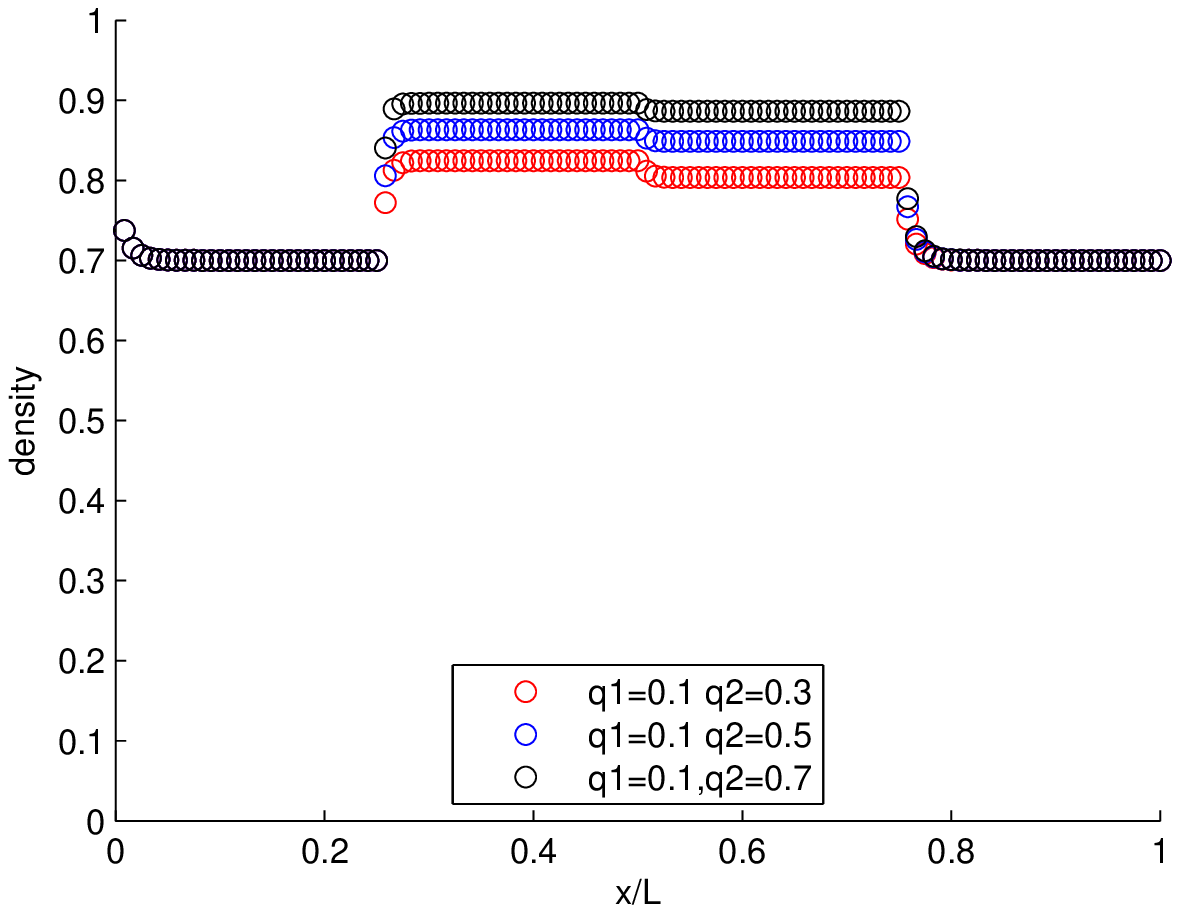}
    \label{fig:side:a}
    \subfigure{\footnotesize (a)}
  \end{minipage}
  \begin{minipage}[t]{0.5\linewidth}
    \centering
    \includegraphics[width=2.5in]{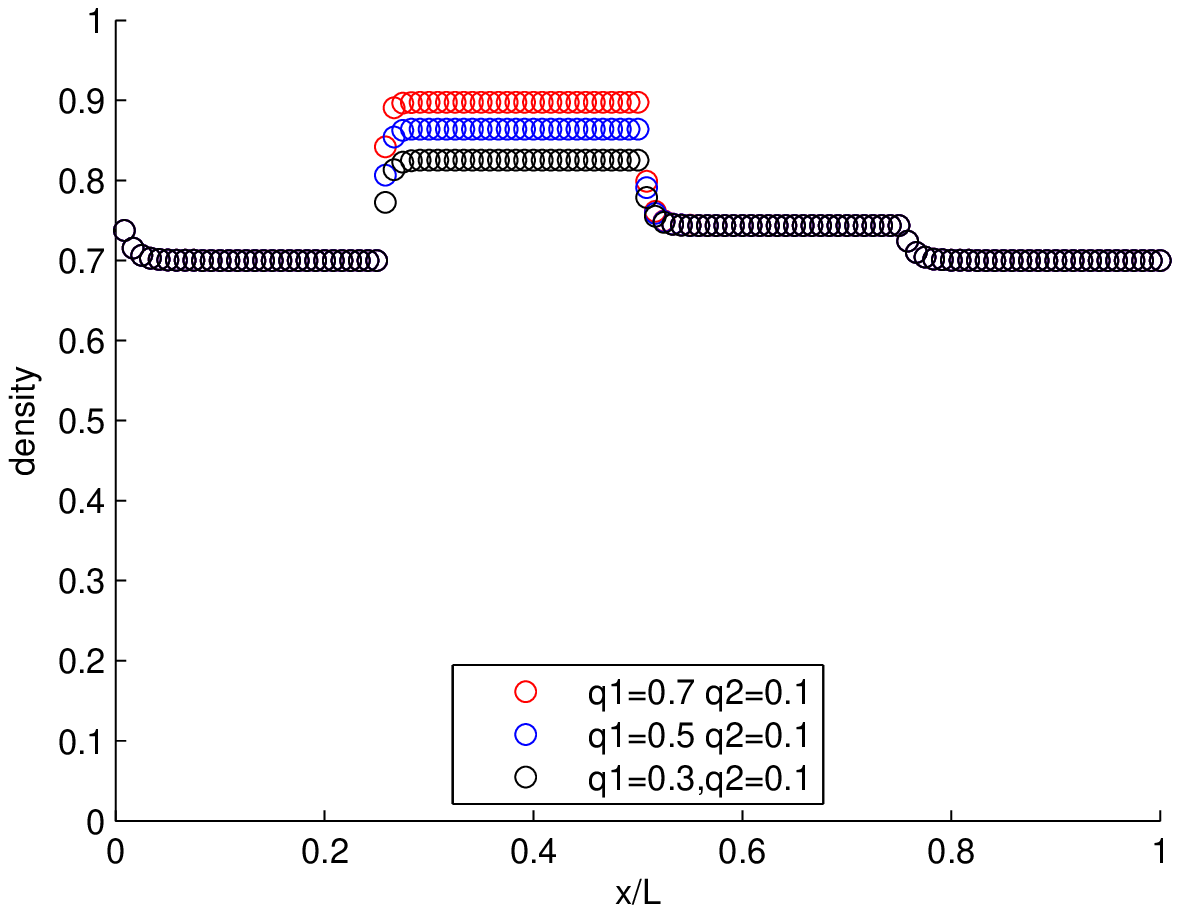}
    \label{fig:side:b}
    \subfigure{\footnotesize (b)}
  \end{minipage}
  \caption{\footnotesize numerical simulations of advanced model 2 with insertion rate $\alpha=0.8$ and
            $\beta=0.3$: (a)$q_{2}$ changes when $q_{1}$ remains constant; (b) $q_{1}$ changes when $q_{2}$ remains constant.}
\end{figure}
\begin{figure}
  \begin{minipage}[t]{0.5\linewidth}
    \centering
    \includegraphics[width=2.5in]{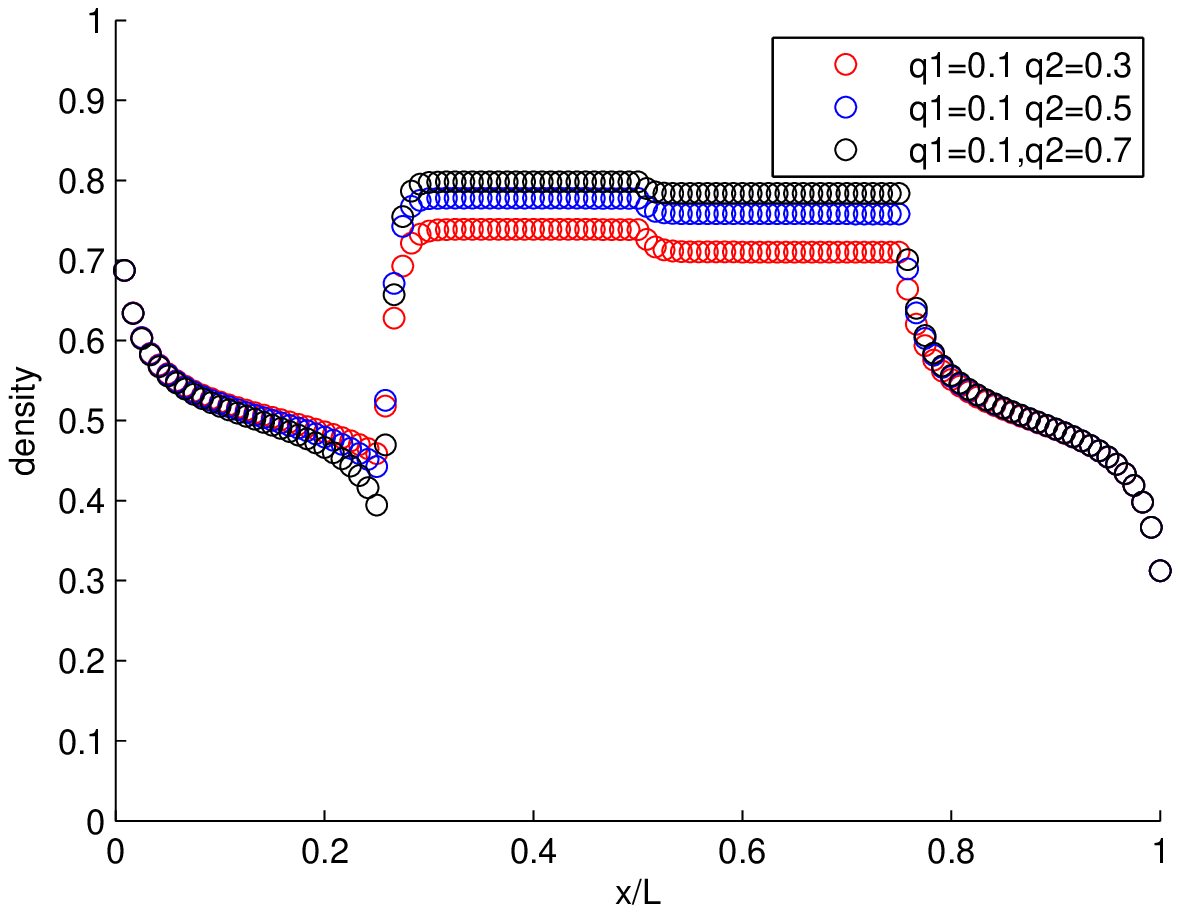}
    \label{fig:side:a}
    \subfigure{\footnotesize (a)}
  \end{minipage}
  \begin{minipage}[t]{0.5\linewidth}
    \centering
    \includegraphics[width=2.5in]{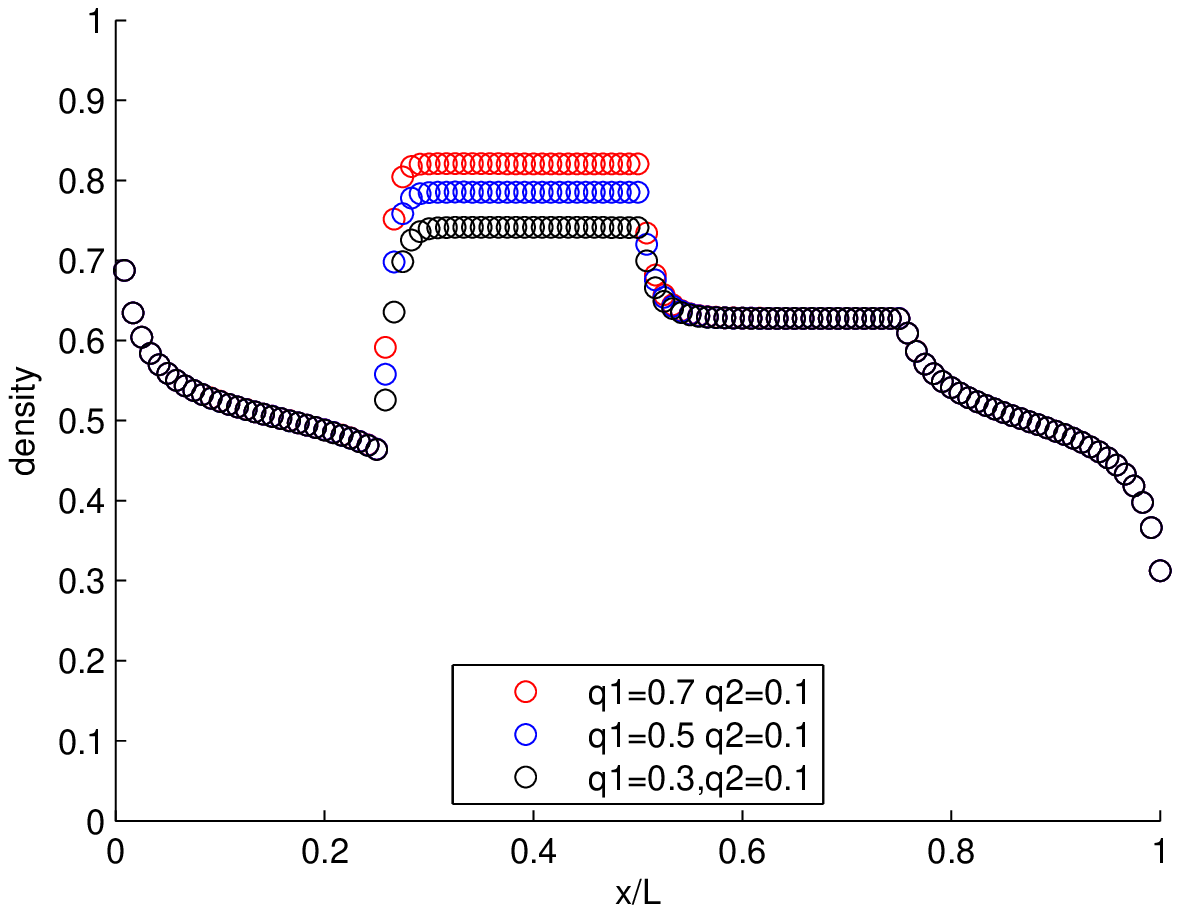}
    \label{fig:side:b}
    \subfigure{\footnotesize (b)}
  \end{minipage}
  \caption{\footnotesize numerical simulations of advanced model 2 with insertion rate $\alpha=0.8$ and
            $\beta=0.8$ with high initial density for each lattice:
           (a)$q_{2}$ changes when $q_{1}$ remains constant; (b) $q_{1}$ changes when $q_{2}$ remains constant.}
\end{figure}
\begin{figure}
  \begin{minipage}[t]{0.5\linewidth}
    \centering
    \includegraphics[width=2.5in]{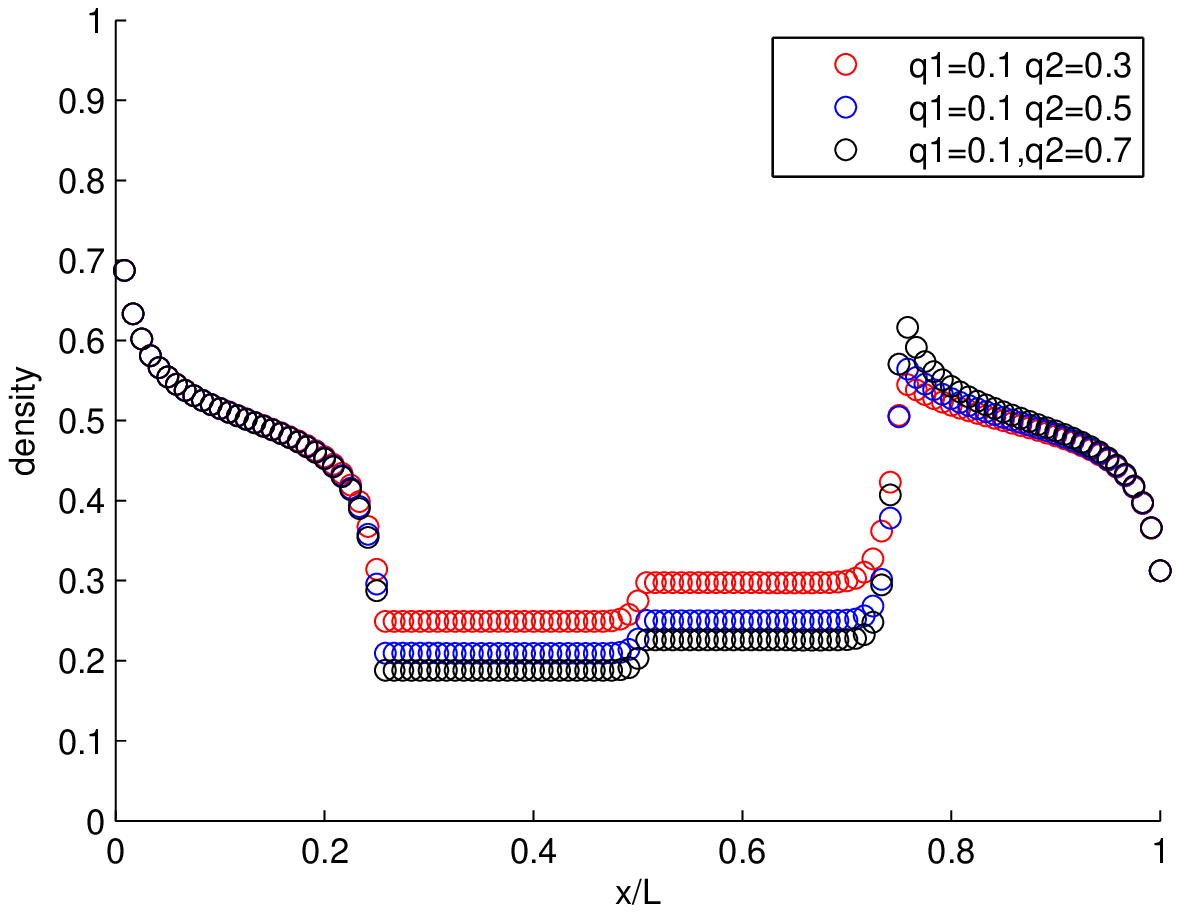}
    \label{fig:side:a}
    \subfigure{\footnotesize (a)}
  \end{minipage}
  \begin{minipage}[t]{0.5\linewidth}
    \centering
    \includegraphics[width=2.5in]{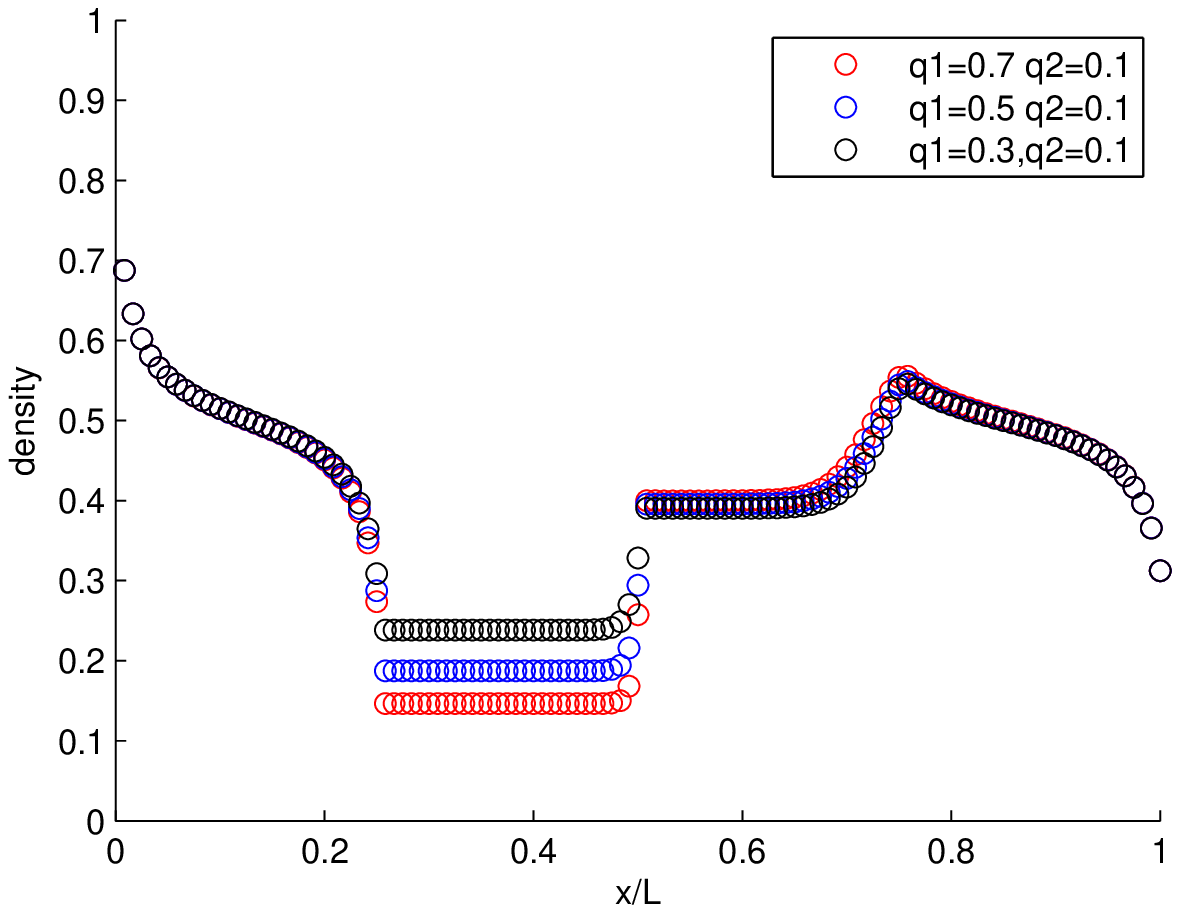}
    \label{fig:side:b}
    \subfigure{\footnotesize (b)}
  \end{minipage}
  \caption{\footnotesize numerical simulations of advanced model 2 with insertion rate $\alpha=0.8$ and
            $\beta=0.8$ with low initial density for each lattice:
           (a)$q_{2}$ changes when $q_{1}$ remains constant; (b) $q_{1}$ changes when $q_{2}$ remains constant.}
\end{figure}
First, from figure 8 to figure 11, we see that the numerical
simulation results are corresponding to our analytical results. The
four segments should be all in low-density phase or high-density
phase, and when segment 1 and 4 are in maximum current phase, the
phase of segment 2 and 3 will be based upon the initial density of
each lattice.

If we focus on figure 8, we will find some interesting things by
comparing figure 8(a) and figure 8(b). In figure 8(a), we let
$q_{2}$ change but $q_{1}$ remain constant. We see that both the
densities in segment 2 and segment 3 decrease in the same pace when
$q_{2}$ increases. In figure 8 (b), we let $q_{1}$ change when
$q_{2}$ remains  constant. The result is that although the density
in segment 2 decreases with increasing of $q_{1}$, the density in
segment 3 does not have apparent change. This phenomenon means that
the change of $q_{1}$ can not influence the density in segment 3
greatly, but the change of $q_{2}$ can really influence the density
in segment 2 greatly. From this we know that the parameter $q_{2}$
has priority to control the density of segment 2 and 3.

From figure 9 to 11, if we compare (a) and (b) of each figure
carefully, it is easy for us to find that for each figure, there
exists the same phenomenon as in figure 8.

In conclusion: in advanced model 2, the density of
segment 2 and 3 depends on parameter $q_{2}$ greatly. The change of
$q_{2}$ will bring about the change of density of both of the
segments. Once $q_{2}$ is set constant, parameter $q_{1}$ will also
has effects on the density in segment 2, but the density in segment
3 will remain in a relatively stable condition.

\section{Conclusion}
In this paper, TASEP with one or two shortcuts have been analyzed in
details. We have found that when a particle arrives at the beginning
of the shortcut, it faces the choice of whether to jump through the
shortcut or not. If it chooses the shortcut, then a problem exists
that whether it can return back to the ordinary road when it finds
the road ahead blocked, according to which we have two different
situations. After the study we know that if the particle can not
return back, the beginning of the shortcut will most probably be
filled with particles, producing a heavy traffic jam. This research
offers us an idea that when we construct a road with a shortcut, the
entrance of the shortcut, that is to say the place where the main
road bifurcates, ought to be built widely. Once a driver chooses to
jump the shortcut but find the signal showing "blocked ahead", he
can immediately turn back to the ordinary road
to prevent time-wasting.

From the simulation of basic model, we find that no matter which
situation is, the three segments should be in the following four
phases:
$$(L,L,L), (H,H,H), (M,L,M), (M,H,M)$$
Whether they are in $(M,L,M)$ or $(M,H,M)$ is based on the initial density
of the lattices.

Based on basic model, two advanced models have been studied. One of
them have two shortcuts at different places while the other have two
shortcuts at the same beginning.

For advanced model 1 we have found that the five segments of the
model should be in the following four phases:
$$(L,L,L,L,L), (H,H,H,H,H), (M,H,M,H,M), (M,L,M,L,M)$$
For advanced model 2 we have found that the four segments of the
model should be in the following four phases:
$$(L,L,L,L), (H,H,H,H), (M,L,L,M), (M,H,H,M)$$

From the simulation of advanced model 1, we have found that this
model can also be regarded as two basic models, the segment 3 of the
first being the segment 1 of the second. The probabilities of the
choices of a particle facing the two shortcuts do not interfere with
each other, which means that the probability of the choice of the
particle facing one of the shortcuts does not influence the density
in the other segment. This phenomenon leads us to suppose that if
there are many shortcuts at different places of the road, we can
view them as a connection of basic models.

From the simulation of advanced model 2, we have found that the
second shortcut can decide the density of both of the two segments
in the middle while the first shortcut can only decide the density
in segment 2. It offers us an idea that if we induce more drivers to
go directly through the second shortcut, the density of the ordinary
road will essentially decrease, providing the whole road is not so
crowded.

However, in this paper all of the proofs are under the condition
$p=1$. Future work can be focused on how to proof the phase
situations  for all $p\in[0,1]$. Also, two shortcuts overlapping
each other will be studied in the future.

\end{document}